\documentclass[prc,showpacs,floatfix,aps]{revtex4}
\usepackage[latin1]{inputenc}
\usepackage{amsmath}
\usepackage{amsfonts}
\usepackage{amssymb}
\usepackage{graphicx}
\usepackage{multirow}

\begin{document}
\title{Periodic Box FHNC calculations of neutron star crustal matter. (I).}
\author{Nicola Bassan}
\email{bassan@sissa.it}
\affiliation{SISSA, I-34136 Trieste, Italy}
\affiliation{INFN, Sezione di Trieste, Trieste, Italy}
\author{Stefano Fantoni}
\email{fantoni@sissa.it}
\thanks{On leave from SISSA, I-34136 Trieste, Italy}
\affiliation{ANVUR, National Agency for the evaluation of Universities and Research Institutes, 20 P.za  Kennedy, I-00144 Roma, Italy}
\author{Kevin E. Schmidt}
\email{kevin.schmidt@asu.edu}
\affiliation{Department of Physics, Arizona State University, Tempe, AZ 85287}
\date{\today}
\begin{abstract}
 Neutron star crustal matter, whose properties are relevant in many models 
aimed at explaining observed astrophysical phenomena, has so far always been 
studied using a mean field approach. In order to check the results obtained in 
this way, a sensible next step is to make use of a realistic nuclear potential. 
The present paper extends the periodic-box Fermi HyperNetted Chain method to 
include longitudinal-isospin dependence of the correlations, making feasible a 
study of asymmetric crustal matter. Results are presented 
for the symmetry energy, the low-density neutron star equation of state and the 
single particle neutron and proton energies.
 \end{abstract}
\maketitle
\section{Introduction}
 Various observed phenomena connected with neutron stars (NS) -- such as glitches -- 
are thought to be largely determined by their crustal 
properties. Glitches are abrupt changes in pulsar periods, \footnote{For a 
statistical study of glitches see \cite{Lyne2000}.} widely thought to be 
related either to starquakes or to vortex pinning, although there are also 
other possibilities \cite{chamel2008}. A key role in these interruptions in 
the usually regular behaviour of the pulsar spin period is thought to be played 
by the interaction between superfluid neutrons and normal matter in the crust. 

Although a number of recent papers have focused on studying NS crustal 
properties (e.g. \cite{Pastore2008} \cite{than2010}), none of them have gone 
beyond a mean field approach for describing the nuclear interaction. That sort 
of approach allows one to give a self-consistent description of all of the 
nucleons forming the crustal matter, either as part of the large neutronized 
neutron-drip nuclei which form the lattice or as part of the neutron superfluid 
which flows through it. (For a recent review of NS crustal matter see 
\cite{chamel2008}). However, major advances have been made in recent years in 
developing nuclear many-body methods for dealing with strongly interacting 
fermions, which have enabled very accurate calculations to be made of several 
properties of nucleonic matter at low temperature, in both normal and 
superfluid phases, which have clearly shown that NN correlations play a 
fundamental role and cannot be disregarded. The short-range behavior of these 
arises from the strong repulsion of the nuclear potential at short 
inter-nucleon distances, and the long-range behavior results from the tensor 
interaction coming from pion-exchange. These types of behavior have proved to 
be a distinctive feature of NN correlations and hence of the nuclear medium, 
and they lead to potentially measurable effects related to NS structure and the 
neutrino mean free path\cite{raffelt1996}.

These technical advances make it possible to perform {\sl ab initio} 
calculations of the structural properties of the NS nuclear medium, all the way 
from the crust to the inner core, fully based on a realistic bare NN 
interaction. This paper is concerned with addressing this challenging problem.

Obviously, {\sl ab initio} calculations have a much more limited range of 
applications than calculations with mean field theories which, however, are 
based on {\sl ad hoc} effective interactions, which are supposed to include the 
main features of the NN correlations in some average way. Therefore, a second 
important goal of these studies is to derive mean field effective interactions 
from first principles, starting from a unique bare nuclear interaction.

We base our investigation here on Orthogonal Correlated Basis theory (OCB) 
\cite{feenberg1969,clark1979,krotscheck1979,fantoni1988} and nuclear Quantum 
Monte Carlo (QMC) methods, particularly Variational Monte Carlo (VMC) and 
Auxiliary Field Diffusion Monte Carlo (AFDMC)\cite{schmidt1999}. All of these 
methods enable one to make very accurate calculations of nuclear matter 
interacting with modern NN potentials, but they have all only been used so far 
for homogeneous matter, for nuclei or for neutron droplets. The NS crustal 
matter is composed of a lattice of neutron-rich nuclei surrounded by superfluid 
neutrons. Typical mixtures are characterized by a lattice length of $\sim 
20-60\,\rm{fm}$ and density values of the neutron soup ranging from $\sim 
0.001\,\rm{fm^{-3}}$ to $\sim 0.1\,\rm{fm^{-3}}$. In spite of having such 
relatively low densities, the neutrons are still in a regime of 
strongly-correlated fermions, because the neutron-neutron scattering length is 
large, $a_{nn}\sim -20 \,\rm{fm}$ and the resulting values of $k_F a_{nn}$ 
range from $\sim -6$ to $\sim -30$.

As far as QMC is concerned, one has to deal with periodic boxes of dimension 
$\sim 20-60 fm$, with an heavy nucleus at the center and a few thousand 
neutrons surrounding it. This requires very resource-intensive simulations, 
which cannot be massively-parallel. Because of this, OCB theory will be used 
first in order to get optimal variational wave functions, and variational 
estimates of the binding energy per particle (as a function of the the neutron 
density and of the symmetry parameter $\alpha=(N-Z)/A$, as well as for 
computing quantities like energy spectra, response functions, spectral 
functions, etc., which are at present beyond the reach of QMC simulations.

In pursuing these goals, we are faced with two kinds of problems: (i) the 
lattice structure of the NS matter does not allow for calculations in 
the thermodynamic limit: systems with $2-3$ thousand neutrons are still far 
away from that; (ii) we are dealing with asymmetric matter having $N>Z$ and we 
cannot just rely on using the two extreme cases $\alpha=0$ and $\alpha=1$ and 
then making a quadratic interpolation between them for all of the other cases, 
because of the presence of the nucleus in the box.

The first of these problems requires relying on the existing PB-FHNC version of 
the FHNC theory. The second one requires us to rewrite the PB-FHNC to allow an 
isospin dependence.

The present paper is devoted to clarifying these two points, which represent 
fundamental methodological steps towards making a truly microscopic and unified 
treatment of NS crustal matter. For doing this, we have extended the PB-FHNC 
method to deal with homogeneous asymmetric nuclear matter described by 
correlated basis functions, in which the correlation operators have a 
longitudinal isospin dependence. Further considerations limit the variational 
choice of the isospin dependence of the correlation operator $\tilde{F}$ 
to its longitudinal component $\tau_z(i)\tau_z(j)$ only. It is then important 
to ascertain how good is such a variational choice compared with the full 
one given by $\mathbf{\tau}(i)\cdot\mathbf{\tau}(j)$. As a first application, we have considered 
simple two-body nuclear potentials of the $v_4$ type, which do not include 
tensor components but have full spin-isospin dependence. Calculations have been 
made of the equation of state (EOS) and of the single particle spectra at 
various values of symmetry parameter.

The results obtained are very encouraging. The iterative process developed for 
solving the new PB-FHNC equations converges rapidly and gives stable solutions. 
Moreover, the longitudinal isospin dependence is able to account for more than 
$80\%$ of the full isospin dependence for the interaction models which we have 
considered, irrespective of the values of the density $\rho$ and the symmetry 
parameter $\alpha$. Interesting results are obtained for the symmetry energy, 
particularly in the low-density region. We compare these results with the 
Bethe-Brueckner-Goldstone (BBG) calculations of ref. \cite{engvik1994}.

Since we want to address this paper and subsequent related ones to the 
astrophysical community, we repeat here some material which has already been 
published in the nuclear physics literature, in order to make the presentation 
comprehensible. We recognise that  readers who are not specialists in nuclear physics 
will need to be strongly motivated in order to work through all of this, but we are aiming 
here to provide a ``bridge'' for those strongly-motivated readers.

The paper is organized as follows. In the next sections we will discuss the 
nuclear interaction, focusing on the two particular interactions that have been 
used in our calculations. Then, in section \ref{sec:FHNC}, we describe the 
state-dependent particle box FHNC scheme used in this work, and finally 
we present our results and give conclusions.

\section{Nuclear potential}

A realistic nuclear potential is usually written as a two-body potential (e.g. 
Argonne V18 \cite{wiringa1995}) plus a three-body contribution (e.g. Urbana IX 
\cite{carlson1983}) which becomes increasingly important beyond half of the 
nuclear saturation density ($\rho_0 = 0.16 \, \rm{fm}^{-3} = 2.7 \times 10^{14} 
\,\rm{g}\,\rm{cm}^{-3}$) (see \cite{lovato2011} and references therein). It 
has been shown that in medium n--body potentials, with $n>2$, can be 
successfully simulated by two--body density dependent terms 
\cite{lagaris1981,lovato2011}.

In this paper, we consider the Illinois class of two--body potentials, which 
are characterized by having a strong local contribution, given by their first 
six spin--isospin--dependent components:
\begin{equation}
\label{eq:v6}
\hat{\upsilon}_{6}(i,j) = \sum_{p=1}^{6} \upsilon^{(p)}(r_{ij}) \hat{O}_{ij}^{(p)} (i,j)\,,
\end{equation}
with:
\begin{equation}
\hat{O}^{p=1-6}_{ij} = (1, \sigma_{ij}, S_{ij} ) \otimes (1, \tau_{ij} )\,,
\end{equation}
 where $\tau_{ij} = \mathbf{\tau}_i \cdot \mathbf{\tau}_j$ and $\sigma_{ij} = 
\mathbf{\sigma}_i \cdot \mathbf{\sigma}_j$, with $\mathbf{\tau}$ and 
$\mathbf{\sigma}$ being the the Pauli matrices acting respectively on the 
isospin and the spin of a nucleon, and $S_{ij} = (3\hat{r}_{ij}^\alpha 
\hat{r}_{ij}^\beta - \delta_{\alpha\beta})\sigma_i^\alpha\sigma_j^\beta$ is the 
tensor operator.

The most important non-local components of the Illinois potentials are the 
spin--orbit terms, $\mathbf{L}_{ij} \cdot \mathbf{S}_{ij}$ and 
$\mathbf{L}_{ij}\cdot \mathbf{S}_{ij}\tau_{ij}$, where $\mathbf{L}_{ij}$ and 
$\mathbf{S}_{ij}$ are the relative angular momentum and the total spin of the 
nucleon pair. Model potentials that include the spin--orbit components are 
denoted as $\upsilon_8$. Other components include $L^2$, 
$(\mathbf{L}_{ij}\cdot \mathbf{S}_{ij})^2$ and symmetry breaking operators, 
giving the total of 18 components of Argonne $\upsilon_{18}$ (AV18) .

In its full form, AV18 gives an almost perfect fit to the NN data up to the 
meson production threshold. Other realistic NN potentials, such as the Bonn and 
Nijmegen potentials, fit the NN data equally well. All of these realistic 
interactions are more or less equivalent in describing the properties of light 
nuclei but the latter ones are basically non-local, and so they are much more 
difficult to handle with many--body theories such as OCB and AFDMC.  Moreover, 
NN correlations are much better described in $r$--space, in which they are 
clearly distinguished from relativistic effects.

There is strong evidence that the first 8 components of the Illinois-type 
potentials are sufficient for giving a realistic description of the nuclear 
medium. Such model potentials can be obtained by simply truncating AV18 after 
the first 8 components. A better choice, however, is to keep the $\upsilon_8$ 
form and re-fit the NN data. That has been done \cite{pudliner1997}, and the 
corresponding model interaction is known in the literature as Argonne 
$\upsilon_{8'}$ or AV8'. The differences between AV18 and AV8' are quite small 
and can safely be treated perturbatively.

Other widely used interaction models from the same class of potentials are 
$\upsilon_{6'}$, which has the form of eq. \ref{eq:v6}, and $\upsilon_{4'}$, in 
which the tensor components are also omitted. The AV6' and AV4' potentials can 
be found in ref.\cite{wiringa2002}. They should really be considered only as 
toy potentials, although they can reproduce a certain amount of nuclear physics 
data. They are certainly very useful, though, for checking many--body 
techniques and for finding the relative importance of the tensor and 
spin--orbit correlations.

\begin{figure}
\includegraphics[width=0.5\textwidth]{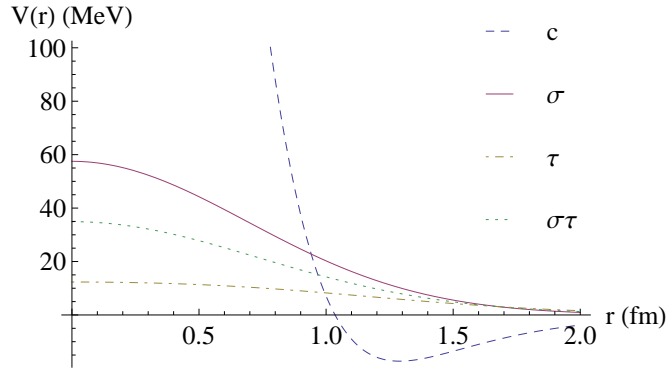}
\caption{\label{fig:diagramATS3} AT4' potential components}
\end{figure}
 One of the potentials which we have used in our calculations is the S3 
potential derived by Afnan and Tang by fitting low energy nucleon-nucleon 
s-wave scattering data (up to 60 MeV) \cite{afnan1968}. This is a $\upsilon_4$ 
potential of the Serber type, and is therefore defined for only the even 
states. It provides a reasonable description of some basic properties of 
$\rm{H^3}$ and $\rm{He^4}$ such as, for instance, the binding energy and the 
root-mean-square radii.  As in the PB--FHNC calculations of ref. 
\cite{fantoni2001c}, we have added to the original S3, an interaction for the 
odd channels given by the repulsive part of the even channels. The four 
components of the resulting potential, which we denote as the AT4' potential, 
are given by
\begin{eqnarray}
\upsilon^{(1)} & \equiv & \upsilon_c = \upsilon_R + \frac{3}{16} (\upsilon_{At}(r_{ij})+\upsilon_{As}(r_{ij}))\, , \nonumber \\
\upsilon^{(2)} & \equiv & \upsilon_\tau =  -\frac{1}{16} (3 \upsilon_{At}(r_{ij}) - \upsilon_{As}(r_{ij}))\, , \nonumber \\
\upsilon^{(3)} & \equiv & \upsilon_\sigma= \frac{1}{16} (\upsilon_{At}(r_{ij}) - 3 \upsilon_{As}(r_{ij}))\, , \nonumber \\
\upsilon^{(4)} & \equiv & \upsilon_{\sigma\tau} = -\frac{1}{16} (\upsilon_{At}(r_{ij})+\upsilon_{As}(r_{ij}))\, , 
\end{eqnarray}
with the repulsive and attractive components given by
\begin{eqnarray}
\upsilon_R &=& \upsilon_{s1} \exp{(-\beta_{s1} r^2)}\, , \nonumber \\
\upsilon_{As} &=& \sum_{i=2}^3 \upsilon_{si} \exp{(-\beta_{si} r^2)}\, , \nonumber \\
\upsilon_{At} &=& \sum_{i=2}^3 \upsilon_{ti} \exp{(-\beta_{ti} r^2)}\, ,
\end{eqnarray}
 where the strengths $\upsilon_{si}$ and $\upsilon_{ti}$ and the gaussian 
coefficients $\beta_{si}$ and $\beta_{ti}$ are given in ref. \cite{afnan1968}. 
The four components of the AT4' potentials are compared in Fig. 1.

 The PB-FHNC calculations of ref. \cite{fantoni2001c} were performed for 
the case of Jastrow correlated theory. The resulting variational energy for a 
system of 2060 nucleons at the experimental equilibrium density $\rho_0$ is 
$-15.150\,\rm{MeV}$, which is not too far away from the FHNC/SOC result, 
$E=-16.184\,\rm{MeV}$ .
 
In order to compare with the Jastrow results of ref. \cite{fantoni2001c}, we have 
also performed variational calculations with the potential $\upsilon_{4}$, 
obtained by truncating $AV8' $ after its first four components, and denoted here as
$(AV8')_4$.  

\section{Correlated Basis Functions\label{sec:FHNC}}

The correlated basis functions for a strongly-correlated Fermi fluid are given 
by
\begin{eqnarray}
\label{eq:varpsi}
\mid n ) &=& \frac{\hat{F}\mid n ]}{[n \mid \hat{F}^\dagger \hat{F}\mid n]^{\frac{1}{2}}} \, ,
\end{eqnarray}
 where $\mid n ] $ is a generic eigenfunction of the Fermi Gas hamiltonian and 
$\hat{F}$ is a correlation operator.  The label $n$ indicates the number of 
particle--hole excitations and $\mid 0 ]$ denotes the Fermi Gas ground state. 
The set of Fermi Gas states $\mid n ]$ is orthonormal, whereas that of the 
correlated states $\mid n )$ is not, because the correlation operator $\hat{F}$ 
breaks the orthogonality. We need to restore orthogonality by following the 
two-step procedure outlined in ref. \cite{fantoni1988}: orthonormal correlated 
states are denoted as $\mid n \rangle$.

\subsection{Properties of the orthonormalization process \label{sec:ortho}}
 It can be shown that the orthonormalization procedure has a number of 
important properties. Formally in OCB theory, the Hamiltonian $H$ is written 
as the sum of an unperturbed term $H_0$ and an interaction term $H_I$:
\begin{eqnarray}
\label{eq:HOCB}
\langle n \mid H_0 \mid m \rangle &=& \langle n \mid H \mid m \rangle \delta_{nm} = H_{nn}\delta_{nm} \, , \nonumber \\
\langle n \mid H_I \mid m \rangle  &=& \langle n \mid H \mid m \rangle (1-\delta_{nm}) = \tilde{H}_{nm} \, .
\end{eqnarray}
 The interaction Hamiltonian $H_I$ is simply the non-diagonal part of $H$. The 
main property of the orthogonalization process of OCB theory is that the 
diagonal matrix elements are not modified by it, i.e.:
\begin{eqnarray}
\label{eq:prop1}
H_{nn} &=& \langle n \mid H \mid n \rangle = (n \mid H \mid n ) + \mbox{terms of order $1/\Omega$}\, .
\end{eqnarray}
 This guarantees that the variational estimates $(I_n \mid H \mid I_n )$ are 
maintained after orthogonalization. A second important property is expressed by 
the following equation:
\begin{eqnarray}
\label{eq:prop2}
(n|H|n) &=& (0|H|0) + \sum_{\mathbf{p}_i}e_v(\mathbf{p}_i) - \sum_{\mathbf{h}_i}e_v(\mathbf{h}_i) 
\nonumber \\
&+& \mbox{terms of order $1/\Omega$}\, .
\end{eqnarray}

In this paper we restrict attention to the diagonal matrix elements of the 
hamiltonian on the ground state $\mid 0 )$ and on the one particle--one hole 
states $\mid \mathbf{p} \mathbf{h} )$, given by
\begin{eqnarray}
\label{eq:DME}
(\mathbf{p} \mathbf{h}| H| \mathbf{p} \mathbf{h}) &=&
\frac{\int dR \mathcal{A}\left[\phi_1 \dots \phi_A \right] \hat{F}_{JL}^\dagger H \hat{F}_{JL} \mathcal{A}\left[\phi_A \dots \phi_1 \right] }
{\int dR \mathcal{A}\left[\phi_1 \dots \phi_A \right] \hat{F}_{JL}^\dagger \hat{F}_{JL} \mathcal{A}\left[\phi_A \dots \phi_1 \right] }\, ,
\end{eqnarray}

where $\phi_n = \exp(i\mathbf{k_n \cdot r_i})\chi_n(s_i)I_n(\tau_i)$ and the 
orbitals $\phi_1 \dots \phi_A$ correspond to the Fermi sea states in the case 
of the ground state energy ($\mathbf{p}= \mathbf{h} = \mathbf{k}_F $) and 
include the excited state orbital $\mathbf{p} $ in place of the Fermi sea state 
orbital $ \mathbf{h}$ in the case of the particle--hole excitation. $\cal{A}$ is
the antisymmetrization operator. Therefore ${\cal A}\left[\phi_1 \dots \phi_A \right] $ 
is a Slater Determinant of plane waves. Note that 
in a Periodic Box treatment, one has a finite number of nucleons and a fixed 
value for the length $L=\Omega^{1/3}$ of the box, given by $\rho_p+\rho_n = 
\rho = A/\Omega$.

Integration over $dR$ extends to all of the $a$ coordinates ($R\equiv \mathbf 
{r}_1, \mathbf {r}_2, \dots, \mathbf {r}_A$) and includes summation over all of 
the spin and isospin variables.

\subsection{The correlation operator}
 Since we are considering a $\upsilon_4$ model interaction, we do not need to 
include tensor correlations in $\hat{F}$. In this case, the standard choice is 
given by
\begin{equation}
\hat{F}_4=\mathcal{S} \left\{ \prod_{i<j=1}^A \hat{f}_4(i,j) \right\} = \mathcal{S} \left\{ \prod_{i<j=1}^A \left[ \sum_{p=1}^4 f^{(p)}(\mathbf{r}_{ij})\hat{O}_{ij}^{(p)}(i,j)\right] \right\} \, ,
\label{eq:corr}
\end{equation}
 where the symmetrization is needed because, in general, the operators 
$\hat{O}_{ij}^{(p)}$ do not commute with each other. In the calculation of the 
matrix elements of any given operator, it is not known how all of the orderings 
of the right hand side of eq. (\ref{eq:corr}) can be taken into account. The best known 
approximation is the so-called FHNC/SOC\cite{pandharipande1979}, which has been 
shown to give reliable results in a number of nuclear matter calculations.  
However, this approximation cannot be used for asymmetric nuclear matter 
because of the presence of the operator $\mathbf{\tau}(i)\cdot\mathbf{\tau}(j)$ 
in $\hat{F}_4$. Because of this, it is important to check the variational 
relevance of the longitudinal isospin-dependent operator 
$\hat{O}_{\tau_z}(i,j)=\tau_z(i)\tau_z(j)$ as compared with 
$\mathbf{\tau}(i)\cdot\mathbf{\tau}(j)$. The operators $\hat{O}_{\tau_z}(i,j)$ 
commute with each other, and so can easily be used in asymmetric nuclear matter 
calculations. To achieve this goal, we consider the following correlation 
operator:
\begin{eqnarray}
\label{eq:FJL}
\hat{F}_{JL} &= &\prod_{i<j=1}\left[ f_{NN}(\mathbf{r}_{ij}) P_{NN}(i,j) +f_{PP}(\mathbf{r}_{ij})P_{PP}(i,j) \right.  
\nonumber \\
		 &+&   \left.  f_{NP}(\mathbf{r}_{ij}) P_{NP}(i,j)  +  f_{PN}(\mathbf{r}_{ij})P_{PN}(i,j)\right]  \, ,
\end{eqnarray}

where the projection operators $P_{ab}$ are given by

\begin{eqnarray}
P_{NN}(i,j) &=& \frac{1 + \tau_z(i)}{2}\frac{1 + \tau_z(j)}{2} \, , \nonumber \\
P_{PP}(i,j) &=& \frac{1 - \tau_z(i)}{2}\frac{1 - \tau_z(j)}{2}  \, , \nonumber \\
P_{NP}(i,j) &=&  \frac{1 + \tau_z(i)}{2}\frac{1 - \tau_z(j)}{2} \, , \nonumber \\
P_{PN}(i,j) &=& \frac{1 - \tau_z(i)}{2}\frac{1 + \tau_z(j)}{2}\, .
\end{eqnarray}

The four scalar functions $f_{NN} $,   $f_{PP} $,  $f_{NP} $ and  $f_{PN} $ have to heal smoothly to 1, thus giving an uncorrelated system, for $r_{ij}$ greater than a certain {\sl healing distance} d chosen so as to minimize the ground state energy per particle of the system $(0|H|0)/A$. The results discussed in this paper are obtained under the following assumption:

\begin{eqnarray}
\label{eq:Fperpar}
f_{NN}(\mathbf{r}_{ij}) &=&  f_{PP}(\mathbf{r}_{ij}) = f_{\parallel}(\mathbf{r}_{ij})\, , \nonumber \\
f_{NP}(\mathbf{r}_{ij}) &=&  f_{PN}(\mathbf{r}_{ij}) = f_{\bot}(\mathbf{r}_{ij})\, .
\end{eqnarray}

 The correlation functions $ f_{\parallel}(\mathbf{r}_{ij})$ and $f_{\bot}(\mathbf{r}_{ij})$ are obtained by  solving a set of second order differential equations \cite{pandharipande1979} whose detailed application to the system at hand is described in appendix \ref{app:corr}. They distinguish isospin parallel from isospin antiparallel correlations.  Such isospin dependence does not completely resolve the difference between 
 $T=1$ and $T=0$ channels as in the case of  $\hat{F}_4$ of eq. (\ref{eq:corr}).  Nevertheless, as will be shown in this paper, it still provides a very 
 good description of the isospin dependence of nuclear correlations induced by a nuclear two--body potential of the $v_4$ type.

\section{Power Series expansion and diagrammatic rules}

The CBF matrix elements given in eq. (\ref{eq:DME}) are calculated by first 
applying standard Fantoni--Rosati (FR) cluster expansion 
techniques\cite{fantoni1975,fantoni1998} to $(\mathbf{p} \mathbf{h}| H 
|\mathbf{p} \mathbf{h}) $ and by summing up the resulting series of cluster 
terms using the FHNC integral equation methods. The FR expansion is based on 
expanding both the numerator and the denominator of eq.(\ref{eq:DME}) in powers 
of the functions $h_\alpha (r)$ given by:
\begin{eqnarray}
h_\alpha (\mathbf{r}) &=& f_\alpha^2(\mathbf{r}) - 1\,, \,\,\,\,  \mbox{($\alpha$ = NN, PP, NP, PN)} \, .
\end{eqnarray}
  
The quantity $\hat{F}_{JL}^\dagger\hat{F}_{JL}$ appearing in the denominator of 
eq. (\ref{eq:DME}) is then decomposed into a series of cluster operators as 
follows:
\begin{eqnarray}
\label{eq:cluster}
\hat{F}_{JL}^\dagger\hat{F}_{JL} &=& \frac{A(A-1)}{2} \hat{f}^2(1,2) \left[1+\sum_{i\neq 1,2}^A \mathcal{X}^{(3)} (1,2;i) + \dots \right] \, ,
\nonumber \\
\hat{f}^2(1,2) &=&  \sum f_\alpha^2 (\mathbf{r}_{12}) P_\alpha(1,2)\,,
\end{eqnarray}
 where each cluster operator $ \hat{f}^2(1,2) \mathcal{X}^{(n)} (1,2;i)$ is 
expressed in terms of products of $h_\alpha$ functions and $\tau^\alpha$ 
projection operators and correlates the two interacting particles $1$ and $2$ 
with the other $n-2$ particles in the medium. A similar expression is obtained for 
the quantity $\hat{F}_{JL}^\dagger \hat{H} \hat{F}_{JL}$ appearing in the 
numerator of eq. (\ref{eq:DME}), where $ \hat{f}^2(1,2)$ is substituted by $ 
\hat{f}(1,2)\hat{H}(1,2)\hat{f}(1,2)$, with $\hat{H}(1,2)$ including the 
potential energy operator $\upsilon_4(1,2)$, and the kinetic energy operators: 
$\nabla^2_1\ \hat{f}(1,2)$, $\nabla_1\hat{f}(1,2)\cdot \nabla_1\hat{ f}(1,i)$, 
$\nabla_1\hat{f}(1,2)\cdot \nabla_1 |n]$, etc.

Inserting the cluster decomposition of eq. (\ref{eq:cluster}) into eq. 
(\ref{eq:DME}), each cluster operator $ \hat{f}^2(1,2)  \mathcal{X}^{(n)} 
(1,2;i)$ gets multiplied by the $n$--body Fermi Gas distribution 
$g_n(1,\dots,n)$. In appendix (\ref{app:standardquantum}), the distribution 
$g_n(1,\dots,n)$ is expressed in terms of the uncorrelated one--body density 
matrix (also called the exchange function)  $ \ell_N(i,j)$  and 
$ \ell_P(i,j)$ which are given by

\begin{eqnarray}
\label{eq:elle}
\ell_N(i,j) &=& \frac{1}{\rho}\sum_{n=1}^N\phi_n^*(i)\phi_n(j) = \ell_N(\mathbf{r}_{ij}) \sum_{m=up,down}\chi^*_m(i)\chi_m(j)
\,, \nonumber \\
\ell_P(i,j) &=& \frac{1}{\rho}\sum_{n=1}^Z\phi_n^*(i)\phi_n(j) = \ell_P(\mathbf{r}_{ij}) \sum_{m=up,down}\chi^*_m(i)\chi_m(j)\, ,
\end{eqnarray}

where the sum is extended over the occupied states. They are normalized to $\rho_N$ and $\rho_P$ respectively. 
It is useful to define a four component  vector function $\mathbf{l}({r}_{ij})$ given by

\begin{eqnarray}
\ell_{NN}(\mathbf{r}_{ij}) &=& \ell_N(\mathbf{r}_{ij})\,, \nonumber \\
\ell_{PP}(\mathbf{r}_{ij}) &=& \ell_P(\mathbf{r}_{ij})\,, \nonumber \\
\ell_{NP}(\mathbf{r}_{ij}) &=& \ell_{PN}(\mathbf{r}_{ij}) = 0  \,. 
\end{eqnarray}

The net result is that   $ \hat{f}^2(1,2)  \mathcal{X}^{(n)} (1,2;i)$  gives rise  to a sum of  n--body {\sl cluster integrals}, whose integrands are products of dynamical correlation functions  $ h_\alpha(\mathbf{r}_{ij})$and exchange correlations  $ \ell_\alpha(\mathbf{r}_{ij})$ .

A very important property of the FR cluster expansion, when applied to finite 
systems, is that in both the numerator and denominator of the diagonal matrix 
element of eq.(\ref{eq:DME}), the summation over the cluster integrals can be 
extended beyond the order A, which is the maximum for a system of A nucleons. 
In fact it can be extended up to infinity because any $g_n(1,\dots,n)$ built with the exchange functions 
$\ell_\alpha(\mathbf{r}_{ij})$ given in eq. (\ref{eq:elle}), with 
$n>A$ vanishes. This property enables us to use all of the FR cluster expansion 
properties which are valid for a system with an unlimited number of nucleons, 
such as for instance nuclear matter.

More details about the FR decomposition can be found in ref. 
\cite{fantoni2001}. We report here only its main properties so as to help the 
reader get a quick understanding of the original papers.

Each cluster integral is most conveniently represented by a {\sl cluster 
diagram}. These diagrams are built by following a few convenient rules 
\cite{fantoni1998}:
   
\begin{enumerate}

\item each point represents a particle. Filled points -- or internal points -- 
represent in-medium particles while unfilled points represent external 
interacting particles.

\item two points are linked either by a dashed line, representing 
$h^\alpha(\mathbf{r})$ or by a solid oriented line, representing 
$\ell(\mathbf{r})$ or by both ;

\item any internal point is reached by one or more dashed lines;  two dashed 
lines cannot be superimposed;

\item solid lines form closed loops and different loops cannot have any common 
point.

\item each internal point carries a proton or neutron density factor, depending 
on which type of particle it represents;

\item all of the particles belonging to an exchange loop are in the same 
spin--isospin state. Each loop (except for those comprising two particles) is 
counted twice, because there is one oriented clockwise and one anti-clockwise. 
The two-particle loops are counted only once. In addition, one has to sum over 
loops of different spin--isospin states. The loop sign is given by $(-)^{n+1}$, 
where n is the number of points on the loop. The global factor $C_n$ of an 
n--particle loop in spin--symmetric matter is given by
\begin{eqnarray}
C_n &=& (-)^{n+1}\times 2\times 2\times (\rho_P^n+\rho_N^n)\left(\frac{1}{2}\right)^n\,,
\end{eqnarray}
 where $\rho_P$ and $\rho_N$ are the proton and neutron matter densities 
respectively. For symmetric nuclear matter ($\rho_P=\rho_N =\rho/2$), the 
factor $C_n$ becomes $ (-)^{n+1}8 (\rho/4)^n$.
 \end{enumerate} 

There are some exceptions to these rules, which occur in the calculation of the 
expectation values of the $\mathbf{\tau}_1\cdot \mathbf{\tau}_2$ and 
$(\mathbf{\sigma}_1\cdot \mathbf{\sigma}_2)(\mathbf{\tau}_1\cdot 
\mathbf{\tau}_2)$ potential terms. These will be discussed later, in 
connection with the calculation of the energy per particle.

The linked cluster theorem\cite{fantoni1998} holds also in the case of the 
$JL$--correlated basis. This theorem states that non-linked cluster diagrams 
(i.e. diagrams which are built from two or more completely unconnected parts 
and which diverge in the thermodynamic limit) cancel exactly between the 
numerator and denominator of eq. \ref{eq:DME} so that one is then left with a 
series of linked cluster diagrams.

\begin{figure}
\includegraphics[width=0.8\textwidth]{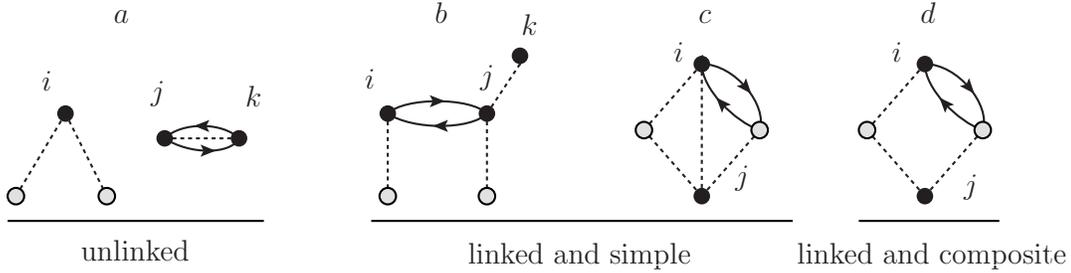}
\caption{\label{fig:diagram1}Examples of cluster diagrams: $A$ is unlinked, $B$ is nodal, $C$ is elementary and $D$ is composite. All the points correspond to neutrons}
\end{figure}

Examples of cluster diagrams are shown in Fig.\ref{fig:diagram1}. Diagram (\ref{fig:diagram1}a) 
is unlinked and is therefore forbidden. The remaining diagrams are all 
allowed. In the calculation of the expectation value of the scalar component 
of the two--body potential diagram, (\ref{fig:diagram1}b) corresponds to the following 
contribution
\begin{eqnarray}
\mbox{Diagram (1b)} &\rightarrow& \frac{1}{2\rho_N}\rho_N^5\int d\mathbf{r}_{12} d\mathbf{r}_i d\mathbf{r}_j d\mathbf{r}_k
f_\parallel^2( \mathbf{r}_{12})\upsilon^c(\mathbf{r}_{12}) \nonumber \\
&\times& h_\parallel(\mathbf{r}_{1i}) h_\parallel(\mathbf{r}_{2j}) h_\parallel(\mathbf{r}_{jk})
\left(\frac{-\ell_N^2(\mathbf{r}_{ij})}{2}\right) \,.
\end{eqnarray}

Linked cluster diagrams are subdivided into two classes: simple and composite 
(or hyper-chain). Simple diagrams are further classified as nodal (or chain) or 
elementary.

Nodal diagrams ($N$--diagrams) are defined as diagrams having one or more {\sl 
nodes}: internal (filled) points that are necessarily crossed by any path going from one 
interacting point to the other. For instance, diagram (\ref{fig:diagram1}b) is nodal and the 
points labelled with $i$ and $j$ are both nodes. Not nodal diagrams
($X$--diagrams) include both the  elementary ($E$--diagrams) and the composite ones.
Composite diagrams are obtained by combining two or more nodal diagrams. Diagram 
(\ref{fig:diagram1}d) is composite and is composed of two nodal subdiagrams.  Elementary 
diagrams ($E$--diagrams) are the remaining ones. They are neither composite nor 
nodal.  They  can be
constructed by first identifying the basic topological structures we want to include. Such basic structures have the property that each of their internal points is reached by at least three links and the two external points by at least two. The links are building blocks given by $N+X$ diagrams. The basic structures are characterized by their number of points, the minimum being four ($E_4$). Diagram (\ref{fig:diagram1}c) is an example of four--point elementary diagrams.

\section{PB--FHNC scheme}

The PB--FHNC integral-equation method\cite{fantoni2001} gives an easy way of 
summing the series of cluster terms by specifying rules for building diagrams, 
using other diagrams as building blocks in an iterative way. The method extends 
Hyper Netted Chain (HNC) theory to the case of correlated quantum Fermi 
systems; HNC theory has been widely used in statistical 
thermodynamics\cite{munster1970} and, more recently, has been applied to CBF 
calculations for low temperature Bose systems such as liquid $^4$He and $^3$He 
impurity in $^4$He\cite{fabrocini1986}. It is based on two basic algorithms: 
the chain and hyper-chain.

\paragraph{\bf Chain algorithm.}  This consists of summing up the whole series of 
$N$--diagrams made from a given building block. One simply takes the sum of 
the geometric series of the Fourier Transforms of the building block function, 
which is given by a subset of $X$--functions

\begin{eqnarray}
{\tilde N}(\mathbf {k}) &=& \rho{\tilde X}^2(\mathbf {k}) +  \rho^2{\tilde X}^3(\mathbf {k}) + \dots =
\frac{ \rho{\tilde X}(\mathbf {k})}{1- \rho{\tilde X}(\mathbf {k})}\,,
\end{eqnarray} 

which can be expressed in terms of the following integral equation:

\begin{eqnarray}
\label{eq:chain}
N(\mathbf{r}_{ij}) &=& \rho \int d\mathbf{r}_{kj} X(\mathbf{r}_{ik})(N(\mathbf{r}_{kj}) + X(\mathbf{r}_{kj})) \,.
\end{eqnarray}

This formula can be interpreted in terms of probabilities. Each integral can be thought of, given two external particles $i,j$, as the probability to find a third in-medium particle $k$ -- expressed by $\rho$ -- times the probability of $i$ interacting with $k$ times the probability of $k$ interacting with $j$. 

\paragraph{\bf Hyper-chain algorithm.} This consists of summing up the whole class 
of $X$--diagrams made from a given subset of $N$--diagrams:

\begin{eqnarray}
\label{eq:hyper}
X(\mathbf{r}_{ij}) &=& f^2(\mathbf{r}_{ij}) \exp{ \left[ N(\mathbf{r}_{ij}) + E(\mathbf{r}_{ij}) \right] } -N(\mathbf{r}_{ij}) - 1\,,
\end{eqnarray}

where $f^2(\mathbf{r}_{ij})$ is given by $\exp (-V(\mathbf{r}_{ij})/KT)$ in statistical thermodynamics and by the correlation 
function of the scalar Jastrow  ansatz $\prod (f(\mathbf{r}_{ij}))$ in the variational calculations of zero temperature Bose systems. The function $E(\mathbf{r}_{ij})$ corresponds to the E-diagrams, which  cannot be calculated in a closed form, like the N-- and X--diagrams. 
The meaning of $X$ becomes clear if we imagine expanding the exponential in series: we are summing an increasing number of $N$ and $E$ diagrams. The result of this procedure will give of course composite and elementary diagrams. In addition to those the exponential term includes  the nodal diagrams and  the identity
which we must   subtract in order to get the sum of the not-nodal diagrams.
The function $E(\mathbf{r}_{ij})$ is a functional of $N(\mathbf{r}_{ij})$ and $X(\mathbf{r}_{ij})$ and 
in general it is approximated by
the first few-body  basic diagrams (the lowest of which is the four--body basic diagram, like the diagrammatic structure underlying
diagram \ref{fig:diagram1}c.

\paragraph{\bf FHNC algorithm}. Eqs. (\ref{eq:chain}) and (\ref{eq:hyper}) are formally identical to the HNC equations 
of statistical thermodynamics and can be solved in an iterative way by means of the following 
steps:

\begin{enumerate}
\item take $N(\mathbf{r}_{ij})=0$;
\item compute $X(\mathbf{r}_{ij})$ using eq. (\ref{eq:hyper}) and taking the function $N(\mathbf{r}_{ij})$ from the previous step;
\item compute  $N(\mathbf{r}_{ij})$ using eq. (\ref{eq:chain}) and taking the functions $X(\mathbf{r}_{ij})$ and  $N(\mathbf{r}_{ij})$ 
on the r.h.s. from the steps 2 and 3 respectively;
\item return to step 2, and continue until convergence is obtained\footnote{If convergence cannot be achieved one can mix the newly computed functions with those computed during the previous iteration.}.
\end{enumerate}

The pair correlation function

\begin{eqnarray}
g(\mathbf{r}_{12}) &=& \frac{A(A-1)}{\rho^2}\ \frac{\int d\mathbf{r}_3 d\mathbf{r}_4\dots|\Psi|^2}
{\int d\mathbf{r}_1d\mathbf{r}_2\dots |\Psi|^2}\,,
\end{eqnarray}

where

\begin{eqnarray}
\label{eq:gjastrow}
g(\mathbf{r}_{12}) &=& 1 +  N(\mathbf{r}_{12}) + X(\mathbf{r}_{12}) = f^2(\mathbf{r}_{12}) \exp(N(\mathbf{r}_{12}) + E(\mathbf{r}_{12}))\,.
\end{eqnarray}

\begin{figure}
\includegraphics{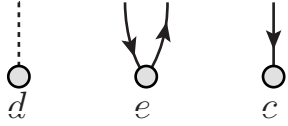}
\caption{\label{fig:diagram2}Labeling of external points. See text for further 
details.}
\end{figure}

In the case of Fermi systems, in addition to the dynamical correlation bonds, 
there are also the exchange bonds, with the diagrammatic rules given in Section 
IV. Because of this, the FHNC method requires further subdivision of diagrams 
by labeling the exchange character of the two external points (e.g. fig 
\ref{fig:diagram2}). Each point in a diagram is labelled with a $d$ unless it 
is reached by an exchange line part of a closed loop (labelled with $e$). 
Therefore one has four different nodal functions $N_{dd}(\mathbf{r}_{ij})$,  $N_{de}(\mathbf{r}_{ij})$,
$N_{ed}(\mathbf{r}_{ij})$ and $N_{ee}(\mathbf{r}_{ij})$, with $N_{de}(\mathbf{r}_{ij}) = N_{ed}(\mathbf{r}_{ji})$. Similarly one has four X-- and
E--functions. In addition one needs to introduce another class of functions, those in which the two external points are joined 
by an open loop of exchange lines, which are denoted with the label $cc$.  This last class of points should not be present in allowed diagrams; diagrams with c points are non-physical (i.e. they have no physical meaning if taken alone) but we include them as useful building blocks. to construct diagram having a closed loop passing through the two external points $1$ and $2$.

It should be noticed that, as in HNC theory, the {\sl external points} $i$ and $j$ of $N$, $X$ and $E$ diagrams 
summed up at a given iteration of the FHNC scheme, may become internal points of the next generation of diagrams. The {\sl true external points}
$1$ and $2$ are those at convergence.

\begin{figure}
\includegraphics[width=0.8\textwidth]{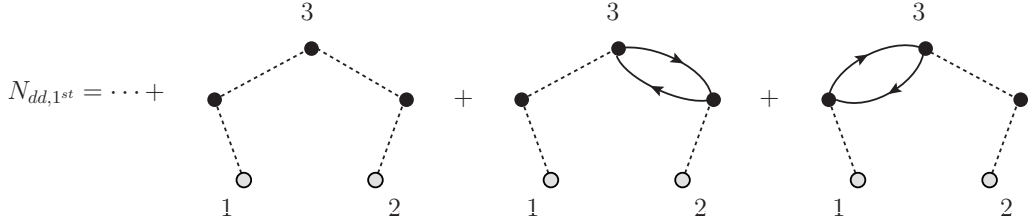}
\caption{\label{fig:diagram3} Building of an $N_{dd}$: graphical counterpart to eq. \ref{eq:ndd}.}
\end{figure}

The chain algorithm for a Fermi system has to take into account the statistical nature of the convolution node, which can be either
$d$ or $e$ in the case of the chain equations for $N_{dd}$, $N_{de}$ and $N_{ee}$, and has to be necessarily
of the type $c$ for $N_{cc}$.
As an example, the equation used to build $N_{dd}(\mathbf{r}_{ij})$   is:
\begin{eqnarray}
N_{dd}(\mathbf{r}_{ij})&=&\rho \int_\Omega d\mathbf{r}_{kj} X_{dd}(\mathbf{r}_{ik}) \left[ X_{dd}(\mathbf{r}_{kj}) + N_{dd}(\mathbf{r}_{kj})\right]  \nonumber \\
&+&\rho \int_\Omega d\mathbf{r}_{kj} X_{dd}(\mathbf{r}_{ik}) \left[ X_{ed}(\mathbf{r}_{kj}) + N_{ed}(\mathbf{r}_{kj})\right]  \nonumber \\
&+& \rho \int_\Omega d\mathbf{r}_{kj} X_{de}(\mathbf{r}_{ik}) \left[ X_{dd}(\mathbf{r}_{kj}) + N_{dd}(\mathbf{r}_{kj})\right] \, ,
\label{eq:ndd}
\end{eqnarray}
where in the first convolution integral the node is of the $d$ type and, in the remaining two, it is of type $e$ . Fig. (\ref{fig:diagram3}) exemplifies 
the above chain equation.

In close analogy with HNC theory, the equation used to build an $X_{dd}$ diagram is
\begin{equation}
X_{dd}(\mathbf{r}_{12})=g_{dd}(\mathbf{r}_{12})-N_{dd}(\mathbf{r}_{12})-1 \, ,
\end{equation}
where 
\begin{equation}
g_{dd}(\mathbf{r}_{12})=f^2(\mathbf{r}_{12})\exp\left(N_{dd}(\mathbf{r}_{12})+E_{dd}(\mathbf{r}_{12})\right) \, .
\end{equation}

The complete set of formulae that are needed to compute the pair distribution function and the energy per particle are given in \cite{fantoni2001}. At present we do not know any formula able to give us a useful prescription to sum $E$ diagrams in a closed form, like those for $N$ and $X$. Luckily enough their contribution has been shown to be almost negligible in the case of translationally invariant nuclear systems \cite{wiringa1988}. This allows us to neglect them for all practical purposes and hence to use the so called FHNC/0 approximation.

\subsection{Vertex corrections}

\begin{figure}
\includegraphics[width=0.8\textwidth] {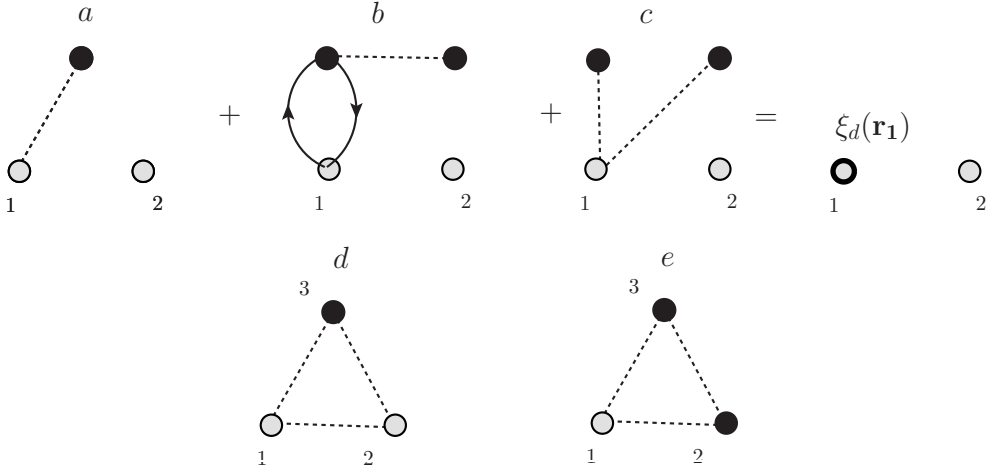}
\caption {\label{fig:diagram5} Diagrammatic exemplification of vertex correction. Diagrams (a-c) are summed up to give $\xi_d(\bf{r}_{1})$. Diagrams (d) and (e) show a basic property of the FHNC equations for $U_d$ and $U_e$. They are the same under integration. However diagram (e) contrary to diagram (d) is invariant under the exchange $P_{23}$. Hence,  it has to be weighted with a prefactor $1/2$. 
That is why in the explicit formula for $U_d$ one has to subtract from $\int d\mathbf{r}_{ij}X_{dd}(\mathbf{r}_{ij})$ 
several other terms that are needed to remove the overcounting coming from these missing prefactors (see ref. \cite{fantoni1998}).}
\end{figure}

Linked diagrams can be divided into two classes: (i) reducible diagrams (like diagram b of fig. \ref{fig:diagram1}) having one or more {\sl reducibility} points, which are the only contact points of two sub-diagrams; (ii) irreducible diagrams  (like diagrams $c$ and $d$ of fig.\ref{fig:diagram1}) which have no reducibility points.
The cluster integral corresponding to reducible diagrams is factorizable in the product of the cluster integrals corresponding to the underlying irreducible diagrammatic structures.
The FHNC scheme sketched in the previous section sums up the irreducible diagrams only. In fact, they are the only remaining diagrams in pure Jastrow theory ($f_\parallel = f_\bot$), because all the reducible ones cancel each other exactly. For instance, diagrams $a$ and $b$ of fig. \ref{fig:diagram5} cancel each other because the exchange loop insertion leads to a factor $-1$ due to
the orthonormality of  the single particle orbitals $\phi_n$.
However the cancellation is no longer true in the more general case of the ansatz given in eq. \ref{eq:corr}  and 
in eq.  \ref{eq:Fperpar} or when the Slater determinant of plane waves is substituted by a BCS wave function to describe a superfluid fermi system, like for instance neutron matter at low density and zero temperature.

It has been proved \cite{fantoni1998} that one can still use the FHNC scheme previously described, paying the price of 
renormalizing  the various points of the irreducible diagrams by proper vertex corrections, which take into account all the 
possible one--body subdiagrams that can be linked to them. The process is exemplified by the diagrammatic equation displayed in fig. \ref{fig:diagram5}. There are two types of vertex corrections, 
 $\xi_d(\mathbf{r})$ for points not reached by exchange lines and  $\xi_e(\mathbf{r})$ for the others .  They are given by the following equations
\begin{eqnarray}
 \xi_d(\mathbf{r})&=&(1+U_e(\mathbf{r}))\exp(U_d(\mathbf{r}))\, , \nonumber \\
 \xi_e(\mathbf{r})&=&\exp(U_d(\mathbf{r})) \, ,
\end{eqnarray} 

where $U_e$ and $U_d$ are the sums of diagrams with $e$ and $d$ starting 
points respectively. In practice, this is accomplished by  integrating  the functions $X_{dd}(\bf{r}_{ij})$ and
$X_{de}(\bf{r}_{ij})$ over  $\bf{r}_j$  for $U_d$ and $X_{ed}(\bf{r}_{ij})$ and
$X_{ee}(\bf{r}_{ij})$ over  $\bf{r}_j$  for $U_e$  and then correcting them to avoid overcounting that would arise because of the increased 
symmetry passing from a two--body diagram to a one--body one. For instance triangular diagrams of  fig. \ref{fig:diagram5}) have
different symmetry factors. The factor is $1$ for the  two--body diagram  (\ref{fig:diagram5}d) and $1/2$  for the one--body diagram  (\ref{fig:diagram5}e). The derivation of the equations leading to  $U_d$ and  $U_e$ can be found in ref. \cite{fantoni1998}

In the above equations $U_d$ comes in the exponential to account for the unlimited number of $U_d$ terms  that can stem from any point
either of type $d$ or type $e$. On the contrary,
one can have  one $U_e$ term only as a vertex correction of a point of type $d$.  It should be noted that  $\xi_d$ corresponds to the
sum of  all the possible one--body linked diagrams and, hence, to the one-body correlation function that, for a homogeneous system, is $g_1(r)=1$. The 
sum rule $\xi_d =1$ can be used as a measure of the accuracy of the FHNC approximation.

As a final remark, we should note that vertex corrected (or renormalized) diagrams do not need to 
obey our third FHNC rule: i.e. there may be diagrams with internal points not 
reached by a dashed line. However this requires that the point should have been 
reached by a dashed line pertaining to a subdiagram accounted for by a 
correction, implying that we will introduce a third correction $\xi_c$ that 
includes the same diagrams as $\xi_e$ except for the identity:

\begin{equation}
\xi_c=\xi_e - 1\, ,
\end{equation}

After having sketched the main  instruments  of PB-FHNC theory we extend it in the following  to include   
longitudinal isospin dependence  in the correlation operator and to the  $N \ne Z$ trial functions.

\section{State dependent PB-FHNC equations}

The  PB--FHNC equations derived in this paper are obtained for the more general form of the 
correlation operator, given in eq. (\ref{eq:FJL}) and not for the restricted  one of eq. (\ref{eq:Fperpar}). 
It follows that nodal, composite, elementary functions, as well as distribution functions, will have the structure of a
four-component vector 

\begin{eqnarray}
\mathbf{A}(\mathbf{r}_{ij}) &=&\left( A^{NN}(\mathbf{r}_{ij}),  A^{PP}(\mathbf{r}_{ij}),   
A^{NP}(\mathbf{r}_{ij}),  A^{PN}(\mathbf{r}_{ij}) \right) \, .
\end{eqnarray}
 where we have also indicated the projectors used in eq. \ref{eq:Fperpar}.

Our nodal equations can be written in a compact way by exploiting the 
convolution formalism. We make the definition:
\begin{equation}
\mathbf{C}_{\alpha}(\mathbf{r}_{ij}) = \left( \mathbf{A}(\mathbf{r}_{ik}) \mid \mathbf{B}(\mathbf{r}_{kj}) \right)_{\alpha} \, ,
\end{equation}
where the subscript $\alpha$ denotes the exchange nature of the node $k$ and, consequently,
 that of the related vertex correction. It can be of type $d$, $e$ or $c$.
 The four components of  $\mathbf{C}$ are given by 
\begin{eqnarray}
 C^{NN}_{\alpha}(\mathbf{r}_{ij}) &=& \sum_{b=N,P}  \rho_{b} \xi_{\alpha}^{b} \int_\Omega d\mathbf{r}_{k} A^{Nb}(\mathbf{r}_{ik})  B^{bN}(\mathbf{r}_{kj})  \, , \nonumber \\
 C^{PN}_{\alpha}(\mathbf{r}_{ij}) &=&  \sum_{b=N,P} \rho_{b} \xi_{\alpha}^{b} \int_\Omega d\mathbf{r}_{k} A^{Pb}(\mathbf{r}_{ik})  B^{bN}(\mathbf{r}_{kj})  \, , \nonumber \\
 C^{NP}_{\alpha}(\mathbf{r}_{ij}) &=&  \sum_{b=N,P} \rho_{b} \xi_{\alpha}^{b} \int_\Omega d\mathbf{r}_{k} A^{Nb}(\mathbf{r}_{ik})  B^{bP}(\mathbf{r}_{kj})  \, , \nonumber \\
 C^{PP}_{\alpha}(\mathbf{r}_{ij}) &=&  \sum_{b=N,P} \rho_{b} \xi_{\alpha}^{b} \int_\Omega d\mathbf{r}_{k} A^{Pb}(\mathbf{r}_{ik})  B^{bP}(\mathbf{r}_{kj})  \, .
\end{eqnarray}

\subsubsection{Nodal diagrams} 
Using the above convolution formalism the chain equations for $\mathbf{N}_{dd}$, $\mathbf{N}_{de}$, 
$\mathbf{N}_{ed}$ and $\mathbf{N}_{ee}$ can be recast in a more compact way, as follows:
\begin{eqnarray}
\mathbf{N}_{dd} &=& \left( \mathbf{X}_{dd} \mid \mathbf{N}_{dd}+ \mathbf{X}_{dd} \right)_{d} +  \left( \mathbf{X}_{de} \mid \mathbf{N}_{dd}+ \mathbf{X}_{dd} \right)_{e} + \left( \mathbf{X}_{dd} \mid \mathbf{N}_{ed}+ \mathbf{X}_{ed} \right)_{e} \nonumber \\
\mathbf{N}_{de} &=& \left( \mathbf{X}_{dd} \mid \mathbf{N}_{de}+ \mathbf{X}_{de} \right)_{d} +  \left( \mathbf{X}_{de} \mid \mathbf{N}_{de}+ \mathbf{X}_{de} \right)_{e} + \left( \mathbf{X}_{dd} \mid \mathbf{N}_{ee}+ \mathbf{X}_{ee} \right)_{e} \nonumber \\
\mathbf{N}_{ed} &=& \left( \mathbf{X}_{ed} \mid \mathbf{N}_{dd}+ \mathbf{X}_{dd} \right)_{d} +  \left( \mathbf{X}_{ee} \mid \mathbf{N}_{dd}+ \mathbf{X}_{dd} \right)_{e} + \left( \mathbf{X}_{ed} \mid \mathbf{N}_{ed}+ \mathbf{X}_{ed} \right)_{e} \nonumber \\
\mathbf{N}_{ee} &=& \left( \mathbf{X}_{ed} \mid \mathbf{N}_{de}+ \mathbf{X}_{de} \right)_{d} +  \left( \mathbf{X}_{ee} \mid \mathbf{N}_{de}+ \mathbf{X}_{de} \right)_{e} + \left( \mathbf{X}_{ed} \mid \mathbf{N}_{ee}+ \mathbf{X}_{ee} \right)_{e}
\end{eqnarray}
The chain equation for $\mathbf{N}_{cc}$ can be written in convolution notation as:
\begin{equation}
\mathbf{N}_{cc}(\mathbf{r}_{12})=\left( \mathbf{X}_{cc} \mid \mathbf{X}_{cc}+ \mathbf{N}_{cc} - \frac{\mathbf{l}}{d} \right)_{e} + 
\left(  - \frac{\mathbf{l}}{d} \mid \mathbf{X}_{cc}+ \mathbf{P} \right)_{e} + \left( - \frac{\mathbf{l}}{d} \mid - 
\frac{\mathbf{l}}{d} + \mathbf{N}_{cc} - \mathbf{P} \right)_{c} \, ,
\end{equation}
where $\mathbf{P}$ is given by:
\begin{equation}
\mathbf{P}(\mathbf{r}_{12})=\left( \mathbf{X}_{cc} \mid \mathbf{X}_{cc}+ \mathbf{N}_{cc} - \frac{\mathbf{l}}{d} \right)_{e} \, .
\end{equation}
These last two equations cannot, however, be applied blindly. According to the 
RF definitions, $cc$ diagrams have exchange-line paths connecting $1$ directly 
to $2$ and so, since exchange correlations cannot flip spins, only 
 $PP$ and $NN$ components are allowed,whereas the components $NP$ and $PN$ vanish  for   $N_{cc}$, $X_{cc}$,$E_{cc}$ and $P$. 
The degeneracy factor
($d$) here is equal to $2$, accounting  only for the two possible spin states ( the isospin states of a pair are singled out in our treatment ). 

\subsubsection{Composite diagrams} 
The equations for the composite functions are a straightforward generalization of the corresponding equations in
standard PB--FHNC theory, and are given by  
\begin{eqnarray}
X_{dd}^\alpha(\mathbf{r}_{12}) &=& g_{dd}^\alpha(\mathbf{r}_{12}) - N_{dd}^\alpha(\mathbf{r}_{12}) - 1 \ , \nonumber \\
X_{de}^\alpha(\mathbf{r}_{12}) &=& X_{ed}^\alpha(\mathbf{r}_{21}) = g_{dd}^\alpha(\mathbf{r}_{12})[N_{de}^\alpha(\mathbf{r}_{12}) 
+ E_{de}^\alpha(\mathbf{r}_{12})] - N_{de}^\alpha(\mathbf{r}_{12})  \ , \nonumber \\ 
X_{ee}^\alpha(\mathbf{r}_{12}) &=& g_{dd}^\alpha(\mathbf{r}_{12})\{N_{ee}^\alpha(\mathbf{r}_{12}) + E_{ee}^\alpha(\mathbf{r}_{12}) 
+ [N_{de}^\alpha(\mathbf{r}_{12}) + E_{de}^\alpha(\mathbf{r}_{12})]^2 \nonumber \\
&-&  d [N_{cc}^\alpha(\mathbf{r}_{12}) - \frac{1}{d}\ell_\alpha(\mathbf{r}_{12}) + E_{cc}^\alpha(\mathbf{r}_{12})]^2 \} - N_{ee}^\alpha(\mathbf{r}_{12})  \ , 
\nonumber \\
X_{cc}^\alpha(\mathbf{r}_{12}) &=& g_{dd}^\alpha(\mathbf{r}_{12}) [N_{cc}^\alpha(\mathbf{r}_{12}) - \frac{1}{2}\ell_a(\mathbf{r}_{12}) 
+ E_{cc}^\alpha(\mathbf{r}_{12})] - N_{cc}^\alpha(\mathbf{r}_{12}) + \frac{1}{2}\ell_\alpha(\mathbf{r}_{12})\ , 
\end{eqnarray}
where the $g_{dd}$ distribution function is defined as
\begin{eqnarray}
g_{dd}^\alpha(\mathbf{r}_{12}) &=&{f^\alpha}^2(\mathbf{r}_{12})\exp[N_{dd}^\alpha(\mathbf{r}_{12}) + E_{dd}^\alpha(\mathbf{r}_{12})]\ .
\end{eqnarray}
It is useful to define also the following four  distribution functions
\begin{eqnarray}
g_{de}^\alpha(\mathbf{r}_{12}) &=& g_{ed}^{\bar\alpha}(\mathbf{r}_{21}) = N_{de}^\alpha(\mathbf{r}_{12}) 
+ X_{de}^\alpha(\mathbf{r}_{12})\ , \nonumber \\ 
(g_{dir})_{ee}^\alpha(\mathbf{r}_{12}) &=& g_{dd}^\alpha(\mathbf{r}_{12})\{N_{ee}^\alpha(\mathbf{r}_{12}) 
+ (E_{dir})_{ee}^\alpha(\mathbf{r}_{12}) + [N_{de}^\alpha(\mathbf{r}_{12}) 
+ E_{de}^\alpha(\mathbf{r}_{12})]^2\} - N_{ee}^\alpha(\mathbf{r}_{12})\ , \nonumber \\
(g_{exch})_{ee}^\alpha(\mathbf{r}_{12}) &=& -d\, g_{dd}^\alpha(\mathbf{r}_{12}) [N_{cc}^\alpha(\mathbf{r}_{12}) 
- \frac{1}{d}\ell_a(\mathbf{r}_{12}) + E_{cc}^\alpha(\mathbf{r}_{12})]^2  
+ g_{dd}^\alpha (E_{exch})_{ee}^\alpha(\mathbf{r}_{12})  \ , \nonumber \\
g_{cc}^\alpha(\mathbf{r}_{12}) &=& N_{cc}^\alpha(\mathbf{r}_{12}) + X_{cc}^\alpha(\mathbf{r}_{12})
 - \frac{1}{d}\ell_\alpha(\mathbf{r}_{12})\ , , 
\label{eq:gfunctions}
\end{eqnarray}
where the components $\alpha=NP$ and $\alpha=PN$ of $(g_{exch})_{ee}$ are identically zero.

\subsubsection{Vertex corrections}  
In deriving the vertex corrections we need to distinguish whether the vertex to be corrected
 is a neutron or a proton. Extending the derivation of ref.\cite{fantoni1998} we get the following expressions.

\begin{eqnarray}
U_d^{N} &=& \sum_{b=N,P} \rho_{b}\int_\Omega d\mathbf{r}_{12}\left\{
\xi_d^{b} [X_{dd}^{Nb}(\mathbf{r}_{12}) - E_{dd}^{Nb}(\mathbf{r}_{12}) - S_{dd}^{Nb}(\mathbf{r}_{12})(g_{dd}^{Nb}(\mathbf{r}_{12})-1)] 
\right.\nonumber \\
&+& \left. \xi_e^{b} [X_{de}^{Nb}(\mathbf{r}_{12}) - E_{de}^{Nb}(\mathbf{r}_{12}) - S_{dd}^{Nb}(\mathbf{r}_{12})g_{de}^{Nb}(\mathbf{r}_{12})
- S_{de}^{Nb}(\mathbf{r}_{12})(g_{dd}^{Nb}(\mathbf{r}_{12})-1)] \right\}+ E_d^{n} \ , 
\end{eqnarray}
for the vertex correction of type $d$, and  
 \begin{eqnarray}
U_{e}^{N} &=& \sum_{b=N,P}  \rho_{b} \int_\Omega d\mathbf{r}_{12}\left\{
\xi_d^{b} [X_{ed}^{Nb}(\mathbf{r}_{12}) - E_{ed}^{Nb}(\mathbf{r}_{12})] + \xi_e^b [X_{ee}^{Nb}(\mathbf{r}_{12}) - E_{ee}^{Nb}(\mathbf{r}_{12})] 
 \right. \nonumber \\
&-& \left. \xi_d^b [S_{dd}^{Nb}(\mathbf{r}_{12})g_{ed}^{Nb}(\mathbf{r}_{12})+S_{ed}^{Nb}(\mathbf{r}_{12})g_{dd}^{Nb}(\mathbf{r}_{12})] 
\right.  \nonumber \\
&-& \left.  \xi_e^b [S_{ee}^{Nb}(\mathbf{r}_{12})(g_{dd}^{Nb}(\mathbf{r}_{12})-1)+  S_{ed}^{Nb}(\mathbf{r}_{12})g_{de}^{Nb}(\mathbf{r}_{12}) + S_{de}^{Nb}(\mathbf{r}_{12})g_{ed}^{Nb}(\mathbf{r}_{12}) +S_{dd}^{Nb}(\mathbf{r}_{12})g_{ee}^{Nb}(\mathbf{r}_{12}) ]\right\}
\nonumber \\
&-& d \rho_N\xi_e^N \int_\Omega d\mathbf{r}_{12} \left\{S_{cc}^{NN}(\mathbf{r}_{12})g_{cc}^{NN}(\mathbf{r}_{12}) 
 + \ell_N(\mathbf{r}_{12})[{\cal N}_{cc}^{NN}(\mathbf{r}_{12}) - \frac{1}{d}\ell_N(\mathbf{r}_{12})]\right\}  + E_{e}^{N}\,, 
\end{eqnarray}
for the vertex correction of type $e$, where $\cal{N}$ is a shorthand for
\begin{equation}
{\cal N}_{cc}^{NN}(\mathbf{r}_{12}) = N_{cc}^{NN}(\mathbf{r}_{12}) 
- \xi_e^{N} \rho_N \int_\Omega d\mathbf{r}_{32} X_{cc}^{NN}(\mathbf{r}_{13})[X_{cc}^{NN}(\mathbf{r}_{32}) + N_{cc}^{NN}(\mathbf{r}_{32})- \frac{1}{d}\ell_N(\mathbf{r}_{32})] \, ,
\end{equation}
and the quantity $S$ is defined as
\begin{equation}
 S_{ij}=\frac{1}{2}\left( N_{ij} + E_{ij}\right)\, .
\end{equation}
Obviously $U^p_{e,d}$ is given by simply interchanging each superscript $p 
\leftrightarrow n$ in the above.

\subsubsection{Two body distribution function} 
The above equations can be solved iteratively. At convergence, the solutions can be used to compute 
the {\sl scalar } two--body distribution function, given by:

\begin{eqnarray}
\label{eq:gscalar}
g_c(\mathbf{r}_{12}) &=&  \sum_{a,b=N,P} g_c^{ab}(\mathbf{r}_{12})
\nonumber \\
&=& \frac{1}{\rho^2} \sum_{a,b=N,P} \rho_a\rho_b \left[ \xi_d^a\xi_d^b g^{ab}_{dd}(\mathbf{r}_{12}) +
 \xi_d^a \xi_e^b  g^{ab}_{de}(\mathbf{r}_{12}) +  \xi_e^a\xi_d^b g^{ab}_{ed}(\mathbf{r}_{12})\right. \nonumber \\
 &+& \left. \xi_e^a\xi_e^b \left(   (g_{dir})_{ee}^{ab}(\mathbf{r}_{12}) + 
  (g_{exch})_{ee}^{ab}(\mathbf{r}_{12})  \right) \right]   \, .
\end{eqnarray}
In the case of symmetric nuclear matter ($N=Z$) and of a state independent correlation operator $(f_\parallel = f_\bot$), 
the quantity $g_c(\mathbf{r}_{12})$ 
recovers the two--body distribution function $g(\mathbf{r}_{12})$ of Jastrow theory  given in eq. (\ref{eq:gjastrow}).  This can be easily understood
by taking into account  (i) the sum rule $\xi_d=1$ and (ii) that $\xi_e g_{de}$ and $\xi_e^2 g_{ee}$ 
of our vertex corrected PB--FHNC theory coincide with the corresponding distribution functions $g_{de}$ and $g_{ee}$ of the standard one.

\subsection{Potential energy expectation value}

The expectation value of a two--body potential of the $v_4$ type on the trial function $\hat{F}_{JL}|0]$ is given by:
\begin{equation}
\frac{\langle V\rangle}{A} = \frac {\rho}{2} \int_\Omega d\mathbf{r}_{12} \{ \upsilon_c(\mathbf{r}_{12}) g_c(\mathbf{r}_{12}) + 3 \upsilon_\sigma(\mathbf{r}_{12}) g_\sigma (\mathbf{r}_{12}) +  \upsilon_\tau (\mathbf{r}_{12}) g_\tau (\mathbf{r}_{12}) + 3 \upsilon_{\sigma \tau} (\mathbf{r}_{12}) g_{\sigma\tau}(\mathbf{r}_{12})\} \, .
\label{eq:potcomponent}
\end{equation}
The first term on the r.h.s.  corresponds to the expectation value of the scalar component of the $v_4$ potential.
We discuss,  in the following, the remaining three terms. 
Recall  that we are dealing with a polarized system with respect to isospin ($N \ne Z$), but with a 
strictly non-polarized one with respect to spin.

\subsubsection{$\sigma_1\cdot \sigma_2$ term} 

The correlation operator $\hat{F}_{JL}$ has no spin dependence. Since $<\sigma_1\cdot\sigma_2> =0$
in spin symmetrical matter, in the calculation of the expectation value of 
$v_\sigma(\mathbf{r}_{12}) \sigma_1\cdot\sigma_2$, the {\sl direct }terms of the distribution function do
not contribute. On the contrary, the exchange terms carry the spin--exchange operator and one has to take care of the following spin algebra

\begin{eqnarray}
<\left(\sigma_1 \cdot \sigma_2 \right)P_\sigma(1,2)> &=& 
<\left(\sigma_1 \cdot \sigma_2 \right) \frac{1+ \sigma_1 \cdot \sigma_2 }{2} >
\nonumber \\
&= & <\frac{3- \sigma_1 \cdot \sigma_2 }{2}>  =   \frac{3}{2}\, ,
\end{eqnarray}
In conclusion we have:
\begin{equation}
g_\sigma (\mathbf{r}_{12}) = \frac{1}{\rho^2}\left( (\xi_e^N \rho_N)^2 
 (g_{exch})_{ee}^{NN}(\mathbf{r}_{12}) 
  +  (\xi_e^P \rho^P)^2  (g_{exch})_{ee}^{PP}(\mathbf{r}_{12})\right)  \, .
\end{equation}

\subsubsection{$\tau_1\cdot \tau_2$ term} 

The operator $\tau_1\cdot\tau_2$ carried by the $\tau$--component of the $\hat{v}_4$ potential
requires a specific and new  PB--FHNC treatment, when dealing with a correlation operator of the $\hat{F}_{JL}$ type and a $N \ne Z$ matter.
We begin by calculating the isospin matrix elements.
The direct terms are: 
\begin{eqnarray}
\langle NN \mid \upsilon_\tau(\tau_1 \cdot \tau_2) \mid nNN\rangle &= &
\langle PP \mid \upsilon_\tau (\tau_1\cdot \tau_2) \mid PP \rangle = \upsilon_\tau  (\mathbf{r}_{12}) \, , \nonumber \\
\langle NP \mid \upsilon_\tau (\tau_1 \cdot \tau_2) \mid NP \rangle &= &
\langle PN \mid \upsilon_\tau (\tau_1 \cdot \tau_2) \mid PN \rangle = - \upsilon_\tau  (\mathbf{r}_{12})\, .
\end{eqnarray}
and the exchange terms are:
\begin{eqnarray}
\langle NN \mid \upsilon_\tau(\tau_1 \cdot \tau_2) P_\tau\mid nNN\rangle &= &
\langle PP \mid \upsilon_\tau (\tau_1\cdot \tau_2) P_tau \mid PP \rangle = \upsilon_\tau  (\mathbf{r}_{12}) \, , \nonumber \\
\langle NP \mid \upsilon_\tau (\tau_1 \cdot \tau_2) P_\tau\mid NP \rangle &= &
\langle PN \mid \upsilon_\tau (\tau_1 \cdot \tau_2) P_\tau\mid PN \rangle = 2 \upsilon_\tau  (\mathbf{r}_{12})\, .
\label{eq:exchangetautau}
\end{eqnarray}
The second row of the above equation  deserves particular attention. Due to the fact that we have e.g. $ \mid NP \rangle$ in the ket and $\langle PN \mid$ in the bra, we have:
\begin{enumerate}
\item a different kind of correlation reaching the \emph{external} point 1 or the external point 2. Under the assumption of eq. (\ref{eq:Fperpar}) there is onyl one type for such new correlations, and we denote it as 
$\delta$--correlation:

\begin{eqnarray}
h_\delta(\mathbf{r}_{ij}) &=& f_\parallel(\mathbf{r}_{ij}) f_\bot(\mathbf{r}_{ij}) -1 \, .
\end{eqnarray}

\item the exchange loop passing through 1 and 2 is made of two cyclic nodal functions one of type $P$ and the other of type $N$.  
\end{enumerate} 
\begin{figure}
\includegraphics{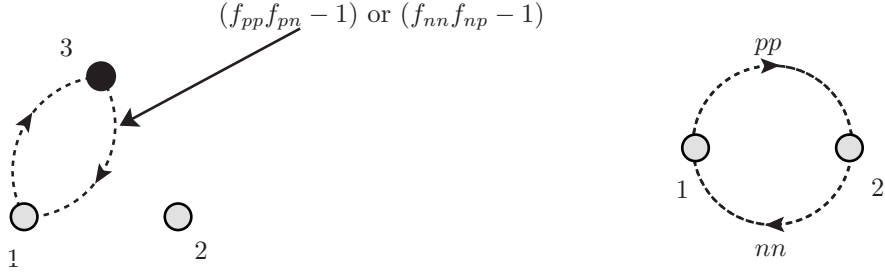}
\caption{\label{fig:diagram4}Left panel: Diagram showing how the new $\delta$ 
correlation results from the isospin-state flip of a particle. Right panel: An 
exchange loop passing through two different external points.}
\end{figure}
It follows that the g-distribution for these matrix elements, which we denote 
by $g_\delta (\vec{r}_{12})$, has to be built by solving appropriate PB--FHNC 
equations, taking into account the above two properties. We have:
\begin{eqnarray}
g_\tau(\mathbf{r}_{12}) &= & \frac{1}{\rho^2}\left( g_\delta(\mathbf{r}_{12}) 
+ \sum_{a=N,P}\left\{\rho_a^2 \left[ (\xi_d^a)^2 
g^{aa}_{dd}(\mathbf{r}_{12}) + \xi_d^a \xi_e^a  \left( g^{aa}_{de}(\mathbf{r}_{12}) 
+  g^{aa}_{ed}(\mathbf{r}_{12}) \right) \right.\right.\right.
\nonumber \\
&+&  \left.\left. \left.(\xi_e^a)^2  \left(   (g_{dir})_{ee}^{aa}(\mathbf{r}_{12}) 
+ (g_{exch})_{ee}^{aa}(\mathbf{r}_{12})  \right) \right] \right\}\right. \\
 &-&\left.\sum_{a=N,P}\rho_a \rho_{\bar{a}} \left[ \xi_d^a \xi_d^{\bar{a}} 
 g^{a{\bar{a}}}_{dd}(\mathbf{r}_{12}) + \xi_d^a \xi_e^{\bar{a}}  g^{a{\bar{a}}}_{de}(\mathbf{r}_{12})
  +\xi_e^a \xi_e^{\bar{a}}  g^{a{\bar{a}}}_{ed}(\mathbf{r}_{12})  
  + \xi_e^a \xi_e^{\bar{a}}  (g_{dir})_{ee}^{a{\bar{a}}}(\mathbf{r}_{12})\right) \right)\nonumber \, ,
\end{eqnarray}
 
where $\bar{a}$ labels the isospin conjugate of $a$, namely 
$\mid\bar{a}\rangle=\tau_x\mid a\rangle$ and the mixed distribution function  $g_\delta$ is given by 
\begin{eqnarray}
g_\delta(\mathbf{r}_{12})  &=& - 4 d\rho_N \rho_P \xi_{e\delta}^2 
 f_\bot^2 e^{N^{NP}_{\delta\delta}(\mathbf{r}_{12})} \left[N^{NN}_{\delta\delta c c}(\mathbf{r}_{12}) 
  - \frac{1}{d} \ell_N(\mathbf{r}_{12}) \right]\left[N^{PP}_{\delta\delta c c}(\mathbf{r}_{12}) 
   - \frac{1}{d} \ell_P(\mathbf{r}_{12}) \right]\, . 
\label{eq:gdelta}
\end{eqnarray}
 
In the case of $N=Z$ matter and $f_\parallel=f_\bot$, one has $\rho_N=\rho_P$, $\xi^N=\chi^P$
with the result that $g_\delta=2g_\sigma$ and  consequently that $g_\tau=3g_\sigma$.

The nodal function  $N^{NP}_{\delta\delta}(\mathbf{r}_{12}) $ is defined by the convolution:
\begin{eqnarray}
N^{NP}_{\delta\delta}(\mathbf{r}_{12}) &=& \left( \mathbf{X}_{\delta d} \mid \mathbf{N}_{d\delta}
+ \mathbf{X}_{d\delta} \right)_{d} +  \left( \mathbf{X}_{\delta e} \mid \mathbf{N}_{d\delta}
+ \mathbf{X}_{d\delta} \right)_{e} + \left( \mathbf{X}_{\delta d} \mid \mathbf{N}_{e\delta}
+ \mathbf{X}_{e\delta} \right)_{e} \, 
\end{eqnarray}
where the nodal and composite vector functions of the type $d\delta$ and $e\delta$  have only the two components  which specify the isospin state of the particle related to the external label
$d$ or $e$. The two chain equations are:
\begin{eqnarray}
\mathbf{N}_{\delta d}(\mathbf{r}_{12}) &=& \left( (\mathbf{X}_{\delta d} \mid
 (\mathbf{N}_{dd}+ (\mathbf{X}_{dd} \right)_{d}
  +  \left( (\mathbf{X}_{\delta e} \mid \mathbf{N}_{dd}+ \mathbf{X}_{dd} \right)_{e} 
  + \left( \mathbf{X}_{\delta d} \mid \mathbf{N}_{ed}+ \mathbf{X}_{ed} \right)_{e}
  \nonumber \\
(\mathbf{N}_{\delta e}(\mathbf{r}_{12}) &=& \left( \mathbf{X}_{\delta d} \mid
 (\mathbf{N}_{de}+ (\mathbf{X}_{de} \right)_{d} 
 +  \left( (\mathbf{X}_{\delta e} \mid (\mathbf{N}_{de}
 + (\mathbf{X}_{de} \right)_{e} + \left( (\mathbf{X}_{\delta d} \mid 
 (\mathbf{N}_{ee}+ (\mathbf{X}_{ee} \right)_{e} \, ,
\end{eqnarray}
and the composite functions are:
\begin{eqnarray}
X_{\delta d}^b(\mathbf{r}_{12}) &=& g_{\delta d}^b(\mathbf{r}_{12})
 - N_{\delta d}^b(\mathbf{r}_{12}) - 1 \ , 
 \nonumber \\
X_{\delta e}^b(\mathbf{r}_{12}) &=& X_{e\delta }^b(\mathbf{r}_{21}) 
= g_{\delta d}^b(\mathbf{r}_{12})[N_{\delta e}^b(\mathbf{r}_{12}) 
+ E_{\delta e}^b(\mathbf{r}_{12})] - N_{\delta e}^b(\mathbf{r}_{12})  \, , 
\end{eqnarray}
with $b=N,P$ and $g_{\delta d}$ given by

\begin{eqnarray}
 g_{\delta d}^b(\mathbf{r}_{12}) &=& f_\parallel(\mathbf{r}_{12}) f_\bot (\mathbf{r}_{12})
 e^{(N_{\delta d}^b(\mathbf{r}_{12})+X_{\delta d}^b(\mathbf{r}_{12}))} \, .
\end{eqnarray}

The two components $N^\alpha_{\delta\delta c c}$, with $\alpha=NN, PP$ , entering 
eq. (\ref{eq:gdelta}), can be given in terms of the following convolution equation
\begin{eqnarray}
\mathbf{N}_{\delta\delta cc}(\mathbf{r}_{12}) &=& \left( \mathbf{X}_{\delta cc} \mid 
\mathbf{X}_{\delta cc}+ \mathbf{N}_{\delta cc} - \frac{\mathbf{l}}{d} \right)_{e} 
+ \left(  - \frac{\mathbf{l}}{d} \mid \mathbf{X}_{\delta cc} +  \hat{P_\delta} \right)_{e}
\nonumber \\
&+ & \left( - \frac{\mathbf{l}}{d} \mid - \frac{\mathbf{l}}{d} 
+ \mathbf{N}_{\delta cc} - \mathbf{P_\delta} \right)_{c} \, ,
\end{eqnarray}
where the two--component vector functions $\mathbf{N}_{\delta\delta cc}$,  
$\mathbf{X}_{\delta\delta cc}$ and $\mathbf{P}_\delta$ are given by

\begin{eqnarray}
\mathbf{N}_{\delta cc}(\mathbf{r}_{12})&=&\left( \mathbf{X}_{\delta cc} 
\mid \mathbf{X}_{cc}+ \mathbf{N}_{cc} - \frac{\mathbf{l}}{d} \right)_{e} 
+ \left(  - \frac{\mathbf{l}}{d} \mid \mathbf{X}_{cc}+ \mathbf{P} \right)_{e} 
\nonumber \\
&+& \left( - \frac{\mathbf{l}}{d} \mid - \frac{\mathbf{l}}{d} + \mathbf{N}_{cc}
 - \mathbf{P} \right)_{c} \, , 
 \nonumber \\
X_{\delta cc}^\alpha(\mathbf{r}_{12}) &=& g_{\delta d}^\alpha(\mathbf{r}_{12}) [N_{\delta cc}^\alpha(\mathbf{r}_{12}) - \frac{1}{2}\ell_a(\mathbf{r}_{12}) + E_{\delta cc}^\alpha(\mathbf{r}_{12})] - N_{\delta cc}^\alpha(\mathbf{r}_{12}) + \frac{1}{2}\ell_a(\mathbf{r}_{12})\, ,
 \nonumber \\
\mathbf{P_\delta}(\mathbf{r}_{12}) &=& \left( \mathbf{X}_{cc} \mid
 \mathbf{X}_{cc\delta}+ \mathbf{N}_{cc\delta} - \frac{\mathbf{l}}{d} \right)_{e} \, .
\end{eqnarray}

In order to complete our PB--FHNC set, we need to define $U^a_\delta$ so as to 
define $\xi^a_{\delta e}$:
\begin{eqnarray}
\xi_{\delta e} &=& e^{U_\delta}\hspace{.5in}\, ,
\\
U_\delta &=& \sum_{b=N,P} \rho^{b}\int_\Omega d\mathbf{r}_{12}\{
\xi_d^{b} [X_{\delta d}^b(\mathbf{r}_{12}) - E_{\delta d}^b(\mathbf{r}_{12}) - S_{\delta d}^b(\mathbf{r}_{12})(g_{\delta d}^b(\mathbf{r}_{12})-1)] 
 \nonumber \\
&+& \xi_e^{b} [X_{\delta e}^b(\mathbf{r}_{12}) - E_{\delta e}^b(\mathbf{r}_{12}) 
- S_{\delta d}^b(\mathbf{r}_{12})g_{\delta e}^b(\mathbf{r}_{12}) 
- S_{\delta e}^b(\mathbf{r}_{12})(g_{\delta d}^b(\mathbf{r}_{12})-1)]\}+E_\delta \, ,
\end{eqnarray}
where:
\begin{eqnarray}
g_{\delta e}^b(\mathbf{r}_{12}) &=& N_{\delta e}^b(\mathbf{r}_{12}) + X_{\delta e}^b(\mathbf{r}_{12})\, .
\nonumber \\
S_{\delta t}^b(\mathbf{r}_{12}) &=& \frac{1}{2} (N_{\delta t}^b(\mathbf{r}_{12})
+E_{\delta t}^b(\mathbf{r}_{12}))\hspace{.5in}\mbox{($t=d,e$)} \, ,
 \end{eqnarray}

\subsubsection{$(\sigma_1\cdot \sigma_2)(\tau_1\cdot \tau_2)$ term}  

The  distribution function  $g_{\sigma\tau}$ can be easily calculated by using the expressions derived in the previoues two subsections.  
\begin{equation}
g_{\sigma\tau} (\mathbf{r}_{12}) =  \frac{1}{\rho^2} \left(\rho_N^2 (\xi_e^n)^2  
(g_{exch})_{ee}^{NN}(\mathbf{r}_{12}) 
  +  \rho_P^2 (\xi_e^p)^2  (g_{exch})_{ee}^{PP}(\mathbf{r}_{12})  + g_\delta\right) \, .
\end{equation}

\subsection{Kinetic energy expectation value}

In this section we calculate the expectation value of the kinetic energy $\langle K \rangle/A$
using the Jackson-Feenberg identity, following the procedure shown in \cite{fantoni1979} 
 and further discussed in \cite{fantoni1998}. 
It is given as  a sum of three terms: a term giving the fermi energy ($E_F$), a term 
accounting for two-body contributions ($E_2$) and a term for three-body 
contributions ($E_3$).
The fermi energy is easily expressed as:
\begin{equation}
E_F = \frac{1}{N}\sum_{filled\,N\,states} \frac{\hbar^2 k_N^2}{2m} + \frac{1}{Z}\sum_{filled\,Z\,states} \frac{\hbar^2 k_P^2}{2m}\, .
\end{equation}
The two body contribution to the kinetic energy can be split into two parts: the first one accounting
for the contribution coming from $ \triangledown^2 $ acting on the correlations 
$f_\parallel$ and $f_\bot$
and the second one accounting for exchanges. The resulting expression for the case of the
$\hat{F}_{JL}$ ansatz and a $N \ne Z$ matter is:
 
\begin{eqnarray}
E_2 &=& E_2^F +E_2^\Phi \, ,
\nonumber \\
E_2^F &=& -\frac{\hbar^2}{4m} \rho\int_\Omega d\mathbf{r} \left[ g^{NN}_c(\mathbf{r}) 
+ g^{PP}_c(\mathbf{r})\right] \triangledown^2 \ln f_\parallel (\mathbf{r}) 
 \nonumber \\
&-&\frac{\hbar^2}{4m}\rho \int_\Omega d\mathbf{r} \left[g^{PN}_c(\mathbf{r})
+g^{NP}_c(\mathbf{r})\right]  \triangledown^2 \ln f_\bot(\mathbf{r})\, , 
\nonumber \\
E_2^\Phi &=& - \frac{\hbar^2}{8m} \frac{\rho_N^2}{\rho} \int_\Omega d\mathbf{r} 
 \left(g_{dd}^{NN}(\mathbf{r})-1\right)  \left(\frac{1}{d}\triangledown^2{\ell_N}^2(\mathbf{r}) 
 - \xi_e^{N} N_{cc}^{NN}(\mathbf{r}) \triangledown^2 {\ell_N}(\mathbf{r}) 
+ 2\xi_c^{N}\frac{\ell_N}{d}\triangledown^2 \ell_N (\mathbf{r}) \right)
  \nonumber \\
&-& \frac{\hbar^2}{8m}  \frac{\rho_P^2}{\rho} \int_\Omega d\mathbf{r} 
 \left(g_{dd}^{PP}(\mathbf{r})-1\right)  \left(\frac{1}{d}\triangledown^2{\ell_P}^2(\mathbf{r})
  - \xi_e^{P} N_{cc}^{PP}(\mathbf{r}) \triangledown^2 {\ell_P}(\mathbf{r})
   + 2\xi_c^{P} \frac{\ell_P}{d}\triangledown^2 \ell_P (\mathbf{r})\right) \, .
\end{eqnarray}
The three body contribution comes from $( \triangledown \ell_{12}\cdot \triangledown \ell_{13})$ and is given by:
 
 \begin{eqnarray}
E_3 &=& - \frac{\hbar^2}{4md}  \sum_{a=N,P}\frac{\rho_a^3}{\rho}
\int_\Omega d\mathbf{r}  \int_\Omega d\mathbf{r'} \triangledown\ell_a (\mathbf{r}) \cdot \triangledown \ell_a(\mathbf{r'}) 
\nonumber \\
&\times&  \left(g_{dd}^{aa}(\mathbf{r})-1\right)\left(g_{dd}^{aa}(\mathbf{r'})-1\right) 
\xi_e^{a} \left[ N_{cc}^a(\mathbf{r-r'}) -\frac{1}{d}\ell_a(\mathbf{r-r'}) 
\right] g_{dd}^{aa}(\mathbf{r-r'}) \, .
\end{eqnarray}

\subsection{Single particle excitation spectrum}

The single particle potential of nuclear matter is calculated by applying the method devised in ref. \cite{friedman1981a} to PB--FHNC 
theory and $N \ne Z$ matter. This will allow us to evaluate the single--particle neutron and proton potentials in neutron rich matter.

It is convenient to calculate, as in ref. \cite{friedman1981a},  the particle--hole excitation energy rather then directly the single particle excitation,
mainly because the  $ \mid\mathbf{p} \mathbf{h})$ state has the same number of particles as the ground state.   Let us consider

\begin{eqnarray}
\epsilon_a(p, h) &=& \langle(\mathbf{p}_a\mathbf{h}_a\mid H \mid \mathbf{p}_a \mathbf{h}_a)
\rangle_{\hat{p}_a,\hat{h}_a} - E_0 \, ,
\end{eqnarray}
 
 where $(\mathbf{p} \mathbf{h}\mid H \mid \mathbf{p} \mathbf{h})$ is defined in eq. (\ref{eq:DME}), the label $a$
 specifies the isospin nature of the excitation, and $\langle\rangle$ stands for average over the directions of $\mathbf{p}$ and
  $\mathbf{h}$. As discussed in section \ref{sec:ortho} the variational estimates of diagonal states are maintained after orthogonalization.
 The single particle excitation $\epsilon_a(q)$ can be obtained  from $\epsilon_a(q, k_F)= \pm\epsilon_a(q) \mp \epsilon_{Fa}$, where
 the upper sign is for particle state ($q>k_F$) and the lower one for hole states ($q<k_F$), and
 
 \begin{eqnarray}
  \epsilon_{Fa} &=& \frac{E_0}{A} + \frac{\rho_a}{A}\frac{\partial E_0}{\partial \rho} \, ,
  \nonumber \\
  &=& \epsilon_0 + \rho_a\frac{\partial \epsilon_0}{\partial \rho} \, .
  \label{eq:epsfa}
 \end{eqnarray}

The single particle spectrum is related to the real part of the nuclear optical potential by

\begin{eqnarray}
\epsilon_a(q) & =&  \frac{\hbar^2q^2}{2m} +U_a(q)\, .
\end{eqnarray}

One can eliminate $q$ from $\epsilon_a(q)$ and $U_a(q)$ to obtain an energy--dependent $U_a(\epsilon)$.
Perturbative corrections  to $\epsilon_a(q)$ and $U_a(q)$  include coupling with two--particle one--hole states
 $\mid \mathbf{p}', \mathbf{p}'', \mathbf{h})$ for $q>k_F$ or two--hole one--particle states 
 $\mid \mathbf{h}', \mathbf{h}'', \mathbf{p})$, for $q<k_F$,  giving a width to  $\mid\mathbf{q})$.  
  
 The particle--hole state $\mid \mathbf{p}_a \mathbf{h}_a)$ is generated in PB--FHNC theory by introducing the following 
 density matrices
 
 \begin{eqnarray}
 l_a(\mathbf{r}_{ij};q_a;k_{Fa}) &=& l_a(\mathbf{r}_{ij}) + \Delta^a(\mathbf{r}_{ij};q_a;k_{Fa})\, ,
 \nonumber \\
 \Delta^a(\mathbf{r}_{ij};q_a;k_{Fa}) &=& \pm \frac{1}{N_{\mathbf{q}_a}}\sum_{n \in \mathbf{q}_a}^ {N_{\mathbf{q}_a}}\phi_n^*(i)\phi_n(j)
		 \mp \frac{1}{N_{\mathbf{k}_{Fa}}}\sum_{n \in \mathbf{k}_{Fa}}^{\mathbf{k}_{Fa}}\phi_n^*(i)\phi_n(j)\, ,
 \end{eqnarray}

where the summations are extended to the  $N_{\mathbf{q}_a}$ states of the shell $\hbar^2 q_a^2/(2m)$ and  
the $N_{\mathbf{k}_{Fa}}$ states of the shell $\hbar^2 k_{Fa}^2/(2m)$.

The cluster diagrams contributing to $\epsilon_a(q, h)$ can be obtained by substituting in all the allowed diagrams 
of $E_0$ one and only one $l_a$--line with a $\Delta^a$--line, for all the $l_a$--lines of the diagram. To sum up the 
resulting series of cluster terms one can use the following algorithm

\begin{description}
\item [1.] modify the density matrices in the following way

\begin{equation}
l_a \rightarrow  l_a + x\Delta^a\, ,
\end{equation}

where $x$ is a smallness parameter and serves to take only one $\Delta$--line at time in each diagram;
\item[2.] solve the PB--FHNC equations with the modified density matrices;
\item[3.] compute the energy expectation value $\epsilon_a(\rho_N,\rho_P;q;k_F;x) $, as for the ground state energy,
with the Fermi energy given by

\begin{equation}
\epsilon_{Fa}(q, k_F) = \frac{\hbar^2}{2m}(\pm q^2 \mp k_{Fa}^2)\, ;
\end{equation}

\item[4.] compute the particle --hole excitation from 

\begin{equation}
\epsilon_a(q, k_F) = \frac{\partial}{\partial x}\epsilon_a(\rho_N,\rho_P;q;k_F;x)\mid_{x=0}\, .
\end{equation}

\end{description}

Note that the discrete character of PB--FHNC  implies that  $\epsilon_{Fa}$ will lie in between two energy shells, that we denote as 
$\epsilon_{Fa-}$ and $\epsilon_{Fa+}$. It follows that the particle energies will be extracted from $\epsilon_a(p_a, k_{Fa-})$ and the
hole states from $\epsilon_a(h_a, k_{Fa+})$. Therefore, one also need to compute $\epsilon_a(k_{Fa+}, k_{Fa-})$.

The neutron and proton effective masses are given by the derivatives

\begin{equation}
\frac{m^*_a(\epsilon)}{m} = 1=\frac{\partial U_a(\epsilon)}{\partial \epsilon}\, .
\end{equation}

Enhancements of $m^*_a(\epsilon)$ will correspond to flattening of $U_a(\epsilon)$ around $\epsilon \sim \epsilon_{Fa}$ which,
most likely, will happen only after having added the perturbative corrections\cite{fantoni1983}.

\section{Results}

In this section we present and discuss {\sl vertex corrected}  PB--FHNC calculations, performed with the
$\hat{F}_{JL}$ model, under the parallel--antiparallel approximation of eq. (\ref{eq:Fperpar})
for the $AT4'$ and $(AV8')_4$ potentials.

We encoded our PB-FHNC scheme as an extension of the code already used in 
\cite{fantoni2001c} and we refer the reader to that paper for details of the 
numerical techniques used. Whenever possible, we made use of standard libraries 
(e.g. fftw3) and routines documented elsewhere (e.g. the ODE integration 
routines from \cite{numrecipes}). We tested our double-precision code with 
different compilers and different optimizations, always obtaining consistent 
results.

\subsection{Comparison of various correlated models}

\begin{table}
\begin{center}
\begin{tabular}{ccccccc}
\hline Potential & Approx. &$A$ & $E_{free}$ & $PE$ & $KE$ & $E$  \\ 
\hline \multirow{3}{*}{$AT4'$}& Jastrow & 2060 & 22.136 &-43.595 & 28.455 & -15.150 \\ 
& $\hat{F}_{JL}$& 2060 & 22.136 & -44.053 & 28.016 & -16.090 \\ 
& $\hat{F}_2$ & $\infty$ & 22.107 & -44.163 & 28.208 & -15.955 \\ 
& $\hat{F}_4$ & $\infty$ & 22.107 & -44.756 & 28.587 & -16.169 \\ 
\hline \multirow{3}{*}{$(AV8')_4$} & Jastrow & 2060 & 22.136 &-27.545& 29.499 & 1.954 \\ 
& $\hat{F}_{JL}$& 2060 & 22.136 & -30.070 & 31.722 & 1.652\\ 
& $\hat{F}_2$& $\infty$ & 22.108 & -28.569 & 30.152 & 1.583 \\ 
& $\hat{F}_4$& $\infty$ & 22.108 & -31.146 & 32.055  & 0.909 \\ 
\hline 
\end{tabular}
 \caption{Results for symmetrical nuclear matter at $\rho=0.16$ for different correlated models. The first
 two rows for each potential are obtained for 2060 nucleons in a periodic box by using
  the vertex corrected PB--FHNC equations. We have used a grid with 60 points in each 
direction.The third and fourth rows report the results obtained
  with FHNC/SOC equations in the thermodynamic limit for the  
  $\hat{F}_2$ and $\hat{F}_4$ models. The energies are in MeV}
 \label{tab:allcomp}
\end{center}
\end{table} 

\begin{figure}
\includegraphics[width=0.6\textwidth, angle=270]{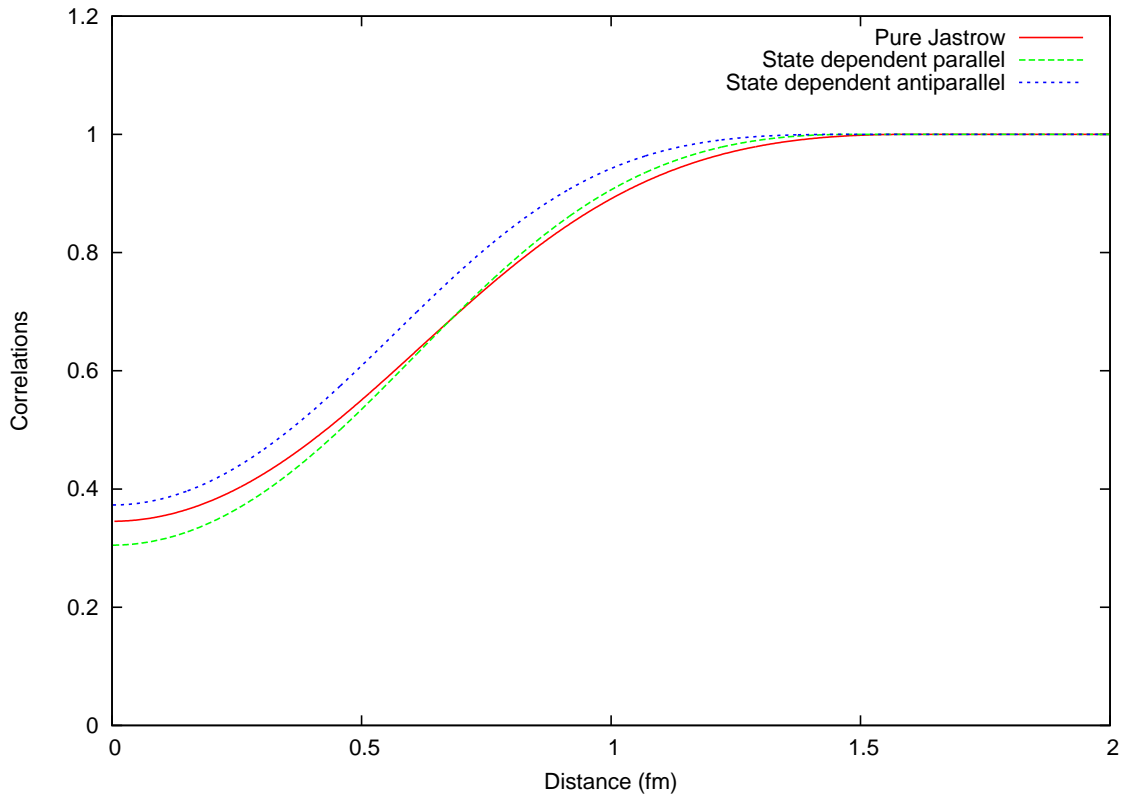}
 \caption{AT4' correlation functions at $\rho=0.16$.  The parallel and antiparallel
 correlations of the $\hat{F}_{JL}$ model are compared with Jastrow correlation.}
\label{fig:correlations}
\end{figure}

We tested our vertex corrected PB-FHNC scheme
 by comparing the results  for SNM with 
the simple Jastrow ansatz with those of ref \cite{fantoni2001c}.
The results obtained for the expectation values of the kinetic energy, $KE$, the potential energy,
 $PE$ and the total energy, $E$, and displayed in the first and the fifth row of Table 
 \ref{tab:allcomp}  coincide within five digits with those of ref \cite{fantoni2001c}.
Note that this check is not at all trivial because it follows from the fulfillment of the sum rule
$\xi_d=1$ and of the relations $\xi_e g_{de} = g_{de}^F$  and 
$\xi_e ^2g_{ee} = g_{ee}^F$ , where $g_{de}^F$ and   $g_{ee}^F$ are the {\sl not vertex corrected}
PB--FHNC distribution functions of ref. \cite{fantoni2001c}.

We have analyzed the quality of our  proposed variational model $\hat{F}_{JL}$ by comparing it
against the results obtained with the $\hat{F}_2$ and $\hat{F}_4$ models  for SNM at $\rho_0$ .
As shown in Table \ref{tab:allcomp}, the longitudinal isospin--dependent model correlation
$\hat{F}_{JL}$ improves considerably the Jastrow ansatz for the $AT4'$ potential. In addition, it
gives equally good energy results as those of  the $\hat{F}_2$ model.  The effectiveness
of   $\tau_z(1)\tau_z(2)$ correlations as compared with the
$\mathbf{\tau}_1\cdot\mathbf{\tau}_2$ ones is confirmed by the results obtained with the $(AV8')_4$
potential which has a stronger $\sigma \cdot \sigma$ dependence.  
 In fig \ref{fig:correlations}, we show our correlation functions computed at 
$\rho=0.16$. As expected from the Pauli exclusion principle, at $r=0$ there is a stronger antiparallel correlation 
and weaker parallel one.

\begin{table}
\begin{center}
\begin{tabular}{cccccc}
\hline $A$ & $E_{free}$ & $PE$ & $KE$ & $E$ & $\Delta_E$ \\ 
\hline 28 & 22.427 & -43.994 & 28.311 & -15.682 & -0.888 \\ 
76 & 21.231 & -44.890 & 27.119 & -17.771 & -0.938 \\ 
108 & 21.277 & -44.933 & 27.163 & -17.769 & -0.932 \\ 
132 & 21.996 & -44.304 & 27.895 & -16.409 & -0.935 \\ 
2060 & 22.136 & -44.053 & 28.016 & -16.090 & -0.940 \\ 
\hline 
\end{tabular}
\caption{PB--FHNC Results for the $AT4'$ potential at $\rho_0$. For the smaller
systems, 26 neighbor cells were summed over; our grid had 60 points in each
direction. $\Delta_E$ gives the difference in energy with respect to the  
Jastrow case. The energies are in MeV}
\label{tab:AT4comp}
\end{center}
\end{table}

\begin{table}
\begin{center}
\begin{tabular}{cccccc}
\hline $A$ & $E_{free}$ & $PE$ & $KE$ & $E$ & $\Delta_E$ \\ 
\hline 28 & 22.427 & -30.937  & 31.971  & 1.034  & -0.270 \\ 
76 & 21.231 &  -30.618 & 30.782 & 0.164 & -0.250 \\ 
108 & 21.277 & -30.590 & 30.826 & 0.235 & -0.249 \\ 
132 & 21.996 & -30.187 &  31.559 & 1.372 & -0.256 \\ 
2060 & 22.136 & -30.070 & 31.722 & 1.652 &  -0.302\\ 
\hline 
\end{tabular}
 \caption{Results for the $(AV8')_4$ potential at 
$\rho_0$. See captions of table \ref{tab:AT4comp}.}
 \label{tab:V4comp}
\end{center}
\end{table} 

Tables \ref{tab:AT4comp} and \ref{tab:V4comp} show the dependence of the energy results on the number
of nucleons in the periodic box. One can see that with $A=132$ finite
size effects are still large. They become totally negligible at $A=2060$ (see also ref.\cite{fantoni2001c})

\subsection{Clustering at subnuclear densities}

\begin{figure}
\includegraphics[width=0.6\textwidth, angle=270]{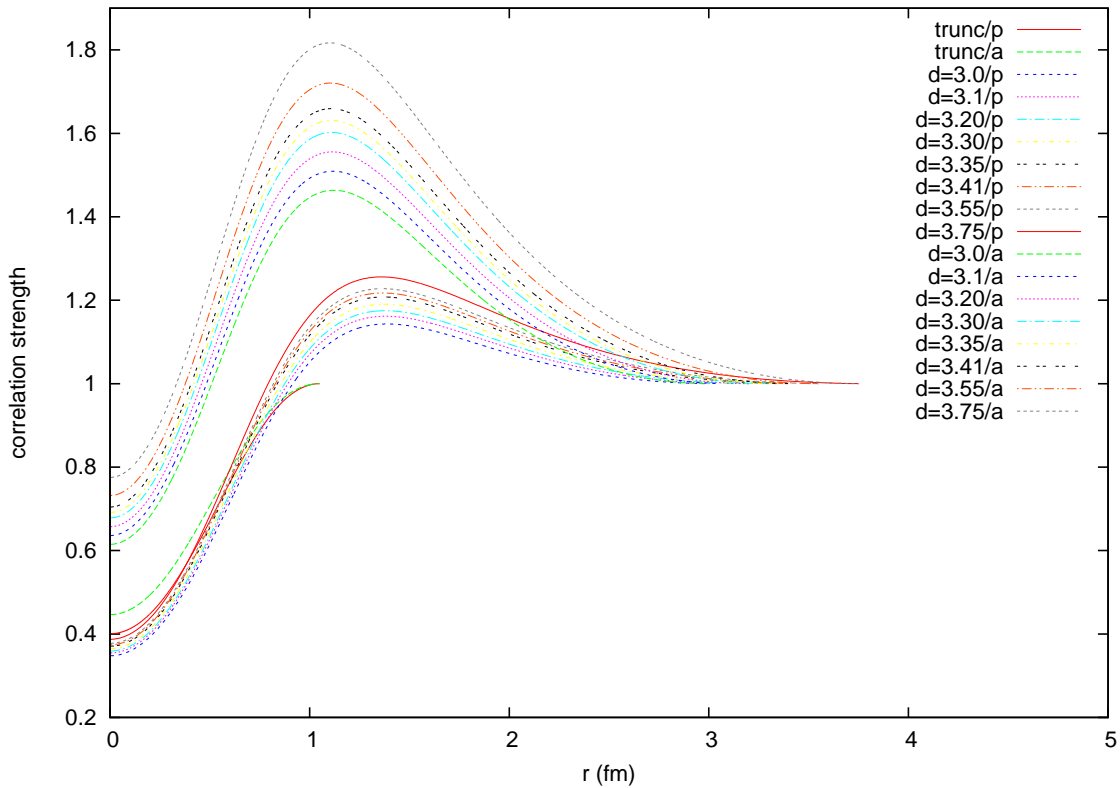}
\caption{AT4' correlation functions at $\rho=0.016$ for different healing distances ($d$) either for $f_\parallel$ (marked with p) or $f_\bot$ (marked with a). We show for comparison two correlation functions whose healing distance was set to 1 so to keep their value everywhere below 1.}
\label{fig:clustering}
\end{figure}

 Although our variational ansatz does not allow an explicit clustering of the nucleonic matter, there are
 strong indications for such clustering phenomena for density below $0.1\rho_0$. To  analyze these
 indications we have studied the behavior of total energy 
  and of the pair distribution functions of SNM at $\rho= 0.1\rho_0$ in two different regions of the healing distance $d$. At
  small $d$ ($d\sim 1\,\rm{fm}$) the correlation functions $f_\parallel$ and $f_\bot$ are below $1$ for $r \leq d$, and the 
  system does not  show any clustering phenomena. On the contrary, as shown in Fig \ref{fig:clustering}, 
  at large $d$ ($3\, \rm{fm}\,\leq d\, \leq 3.35\, {\rm fm}$) 
  the correlation functions   have pronounced peaks, occurring roughly at the same value of the interparticle distance, 
  irrespective of the value of   $d$, as typically happens in clustering phenomena.  One can see from 
  Table \ref{tab:clusteringE} that the variational energy gets lower in the region of large $d$ reaching a 
  minimum around $d\sim 3.3\,\rm{fm}$. 
 
\begin{figure}
\includegraphics[width=0.7\textwidth, angle=270]{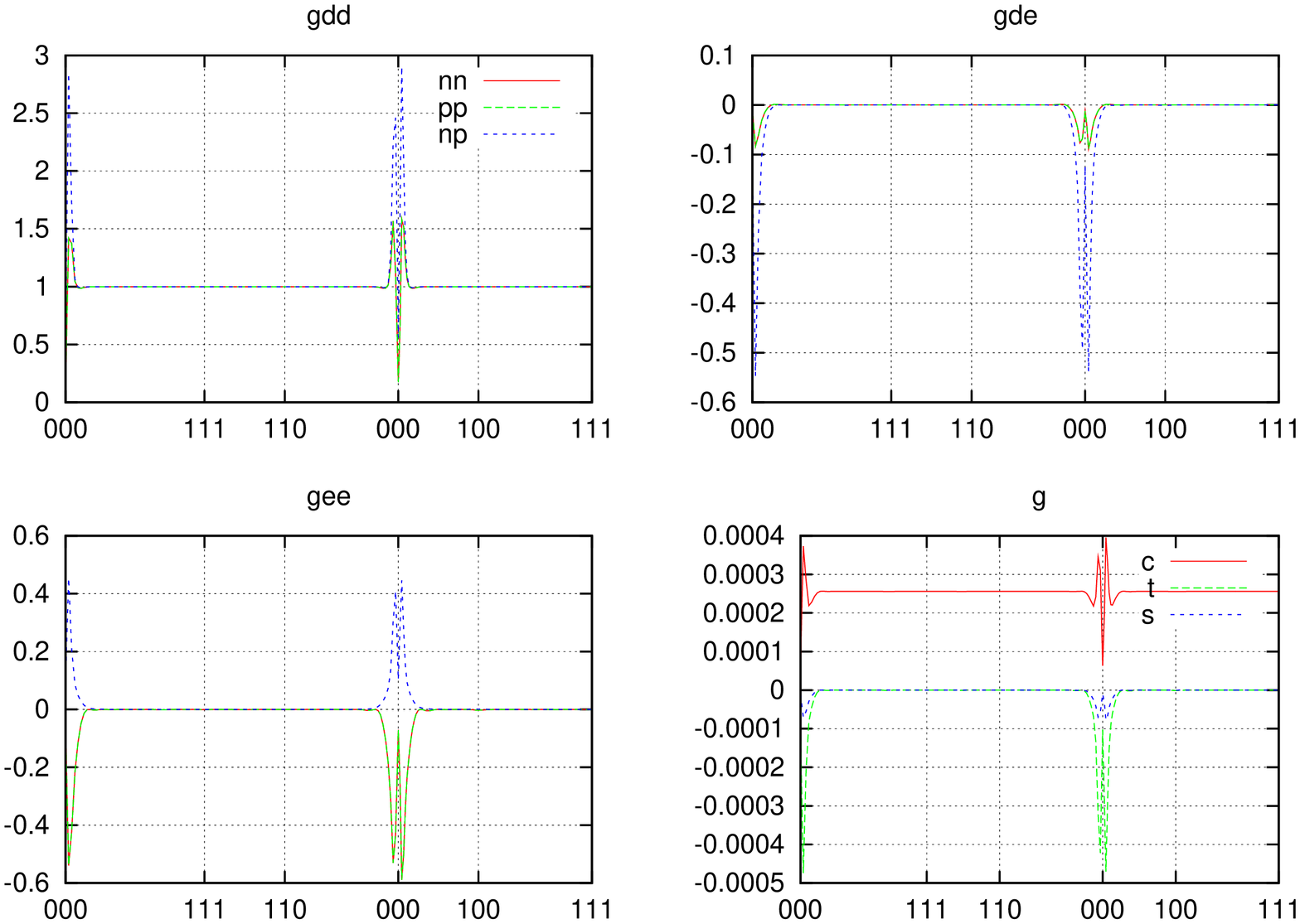}
\caption{AT4' gfunctions at $\rho=0.016$ for d=3.30.}
\label{fig:gclustering}
\end{figure}

In table \ref{tab:clusteringE} we show the energy of a system of 2060 nucleons (SNM) computed with different healing distances. For all these calculations we enforced the same level of convergence of our FHNC equations. For $d>3.35\,\rm{fm}$ we were not able to make our equations converge any longer. For comparison we show the energy obtained with $d=1\,\rm{fm}$ which is the largest healing length that does not produce the clustering effect.

In figure \ref{fig:gclustering} we show our results, obtained with $d=3.30 \rm{fm}$ for the various $g$ components and for $g_{dd},\, g_{de} \rm{and}\, g_{ee}$. It is easy to see that indeed in the channel $np$ there is a clear evidence of clustering occuring.

\begin{table}
\begin{center}
\begin{tabular}{c|cccccc}
\hline $d\,\rm{fm}$  &1 &  3        &   3.1    &  3.2      &  3.3     &  3.35   \\ 
\hline $E\,\rm{MeV}$  &-0.73 & -2.634 & -2.787 & -2.950 & -3.087 & -3.097 \\ 
\end{tabular}
\caption{SNM energy per particle computed at $\rho=0.016$ using the AT4' potential.  Our grid was set up to have 45 points in every direction.}
\label{tab:clusteringE}
\end{center}
\end{table} 

Our results strongly indicate that at such low densities variational functions allowing for a nucleus embedded in a neutron fluid would have a lower energy with respect to those describing an homogeneous fluid. In the following we have always forced the system to behave as an homogeneous fluid.

\subsection{Symmetry energy at subnuclear densities}

One of the main advantages of a $\tau_{z1}\tau_{z2}$ form of the isospin correlations is that one can easily compute the energy expectation value of  nuclear matter with $N \neq Z$, provided that both $N$ and $Z$ are magic numbers (i.e. they correspond to shell closure). For this reason, we cannot keep $A=N+Z$ fixed. The average value of A which we find more convenient because it is sufficiently large to reduce finite size effects and allows for a quite large number of admixtures with the smallest fluctuations ($\Delta A$) is the magic number $1898$. In table \ref{tab:magic} we report the number of protons and neutrons for each admixture, the percentage $Z/A=0.5(1-\alpha)$ and the energy results of our calculations - performed using AT4' potential - at 4 different densities. 

\begin{table}
\begin{center}
\begin{tabular}{cccccccccc}
  $Z$  & $N$ &  $A$   &   $\Delta A$    &  $\%$   &  $\alpha^2$  & $E(10^{-2}\rho_0)$& $E(10^{-1}\rho_0)$ & $E(\rho_0)$ & $E(\rho=0.25)$\\ 
\hline 
0	&1898	&1898	&0	&0,00\%	 &1,000	& 1.26	& 4.94	& 25.55	& 57.13\\
54	&1850	&1904	&6	&2,84\%	 &0,890	& 1.16	& 4.28	& 20.78	& 47.52\\
114	&1790	&1904	&6	&5,99\%	 &0,775	& 1.05	& 3.62	& 15.95	& 39.30\\
186	&1694	&1880	&-18&9,89\%	 &0,643	& 0.93	& 2.86	& 10.44	& 29.00\\
294	&1598	&1892	&-6	&15,54\% &0,475	& 0.78	& 1.91	&  3.44	& 15.84\\
406	&1502	&1908	&10	&21,28\% &0,330	& 0.65	& 1.08	& -2.64	& 4.40 \\
514	&1382	&1896	&-2	&27,11\% &0,210	& 0.55	& 0.42	& -7.55	& -4.88\\
730	&1174	&1904	&6	&38,34\% &0,054	& 0.42	& -0.43	& -13.84& -16.79\\
874	&1030	&1904	&6	&45,90\% &0,007	& 0.38	& -0.70	& -15.88& -20.67\\
\end{tabular}
\caption{Number of protons and neutrons for different admixtures that we studied and results at different densities.}
\label{tab:magic}
\end{center}
\end{table} 

We used our results to check whether the standard way of fitting asymmetric admixture energies (i.e. using a simple quadratic fit) can be reliably used at sub-nuclear densities. We used Mathematica to fit our results using a 6th degree polynomial as a prior. As expected we get null coefficients for the odd power terms. We also get non zero coefficients for the 4th power term. Such coefficients however are always negligible, being, at most, one order of magnitude smaller than the 2nd power term one. 

\begin{figure}
\includegraphics[width=0.6\textwidth]{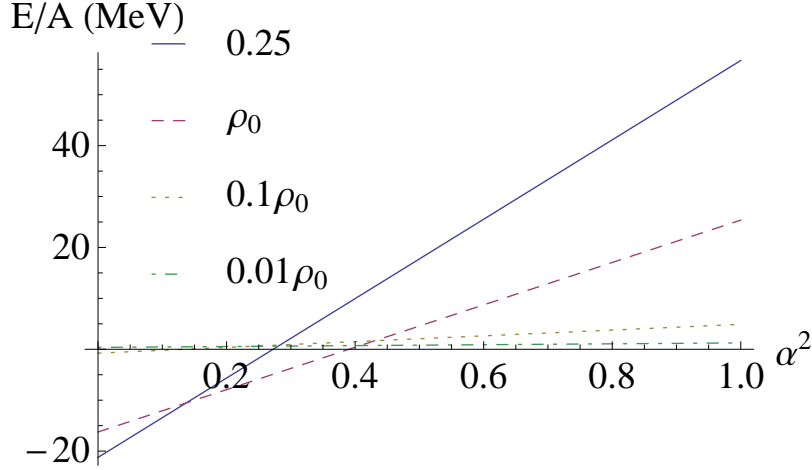}
\caption{Energies per particle ($E/A$) of admixtures with different $\alpha^2$ at different densities.}
\label{fig:asymm}
\end{figure}

In figure \ref{fig:asymm} we show our results and our fits for the energies ($E$) of admixtures with different $\alpha^2$. In table \ref{tab:symmE} we give our best estimate for the symmetry energy ($S$) at different densities computed using the AT4' potential and a purely quadratic fit. The value obtained at $\rho_0$ is somewhat larger than the experimental results of $S \sim 36\,\rm{MeV}$; this is expected due to the phenomenological nature of our potential.

\begin{table}
\begin{center}
\begin{tabular}{c|cccc}
\hline $\rho\,(\rm{fm^{-3}})$  &$10^{-2}\rho_0$& $10^{-1}\rho_0$ & $\rho_0$ & $0.25$   \\ 
\hline $S \, (\rm{MeV})$  & 0.89 &  5.66 & 41.62& 77.96 \\ 
\end{tabular}
\caption{Symmetry energy computed at different densities using the AT4' potential.}
\label{tab:symmE}
\end{center}
\end{table}

\subsection{Equations of state for AT4'}
\begin{figure}
\includegraphics[width=0.6\textwidth]{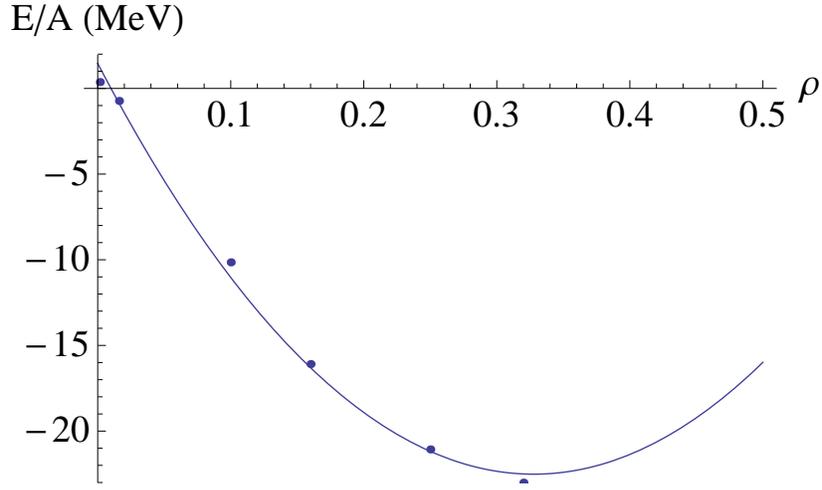}
\caption{Energy per particle of SNM computed at different densities using the AT4' potential. For low densities we forced the system to be homogeneous.}
\label{fig:snm_e}
\end{figure}

In figure \ref{fig:snm_e} we show our results for the energy of SNM computed at different densities using the AT4' potential. We find that we can nicely fit our data by:

\begin{equation}
E_{SNM}(\rho)=E_0+a(\rho-\rho_{eq})^2+b(\rho-\rho_{eq})^3 e^{\gamma(\rho-\rho_{eq})}\,
\end{equation}

where $E_0=-22.51 \,\rm{MeV}$, $\rho_{eq}=0.33\,\rm{fm^{-3}}$, $a=220\,\rm{MeV\,fm^6}$, $b=-1.56\,\rm{MeV\,fm^9}$ and $\gamma = -5.570\,\rm{fm^3}$. The EOS for asymmetric matter can then be written, fitting the data for different $\alpha$, as:

\begin{equation}
E(\rho,\alpha)=E_{SNM}(\rho)+C_s\left(\frac{\rho}{\rho_{eq}}\right)^{\gamma_s} \alpha^2\,
\end{equation}

where $C_s=82.86\,\rm{MeV}$ and $\gamma_s=0.913$.

\begin{figure}
\includegraphics[width=0.6\textwidth, angle=270]{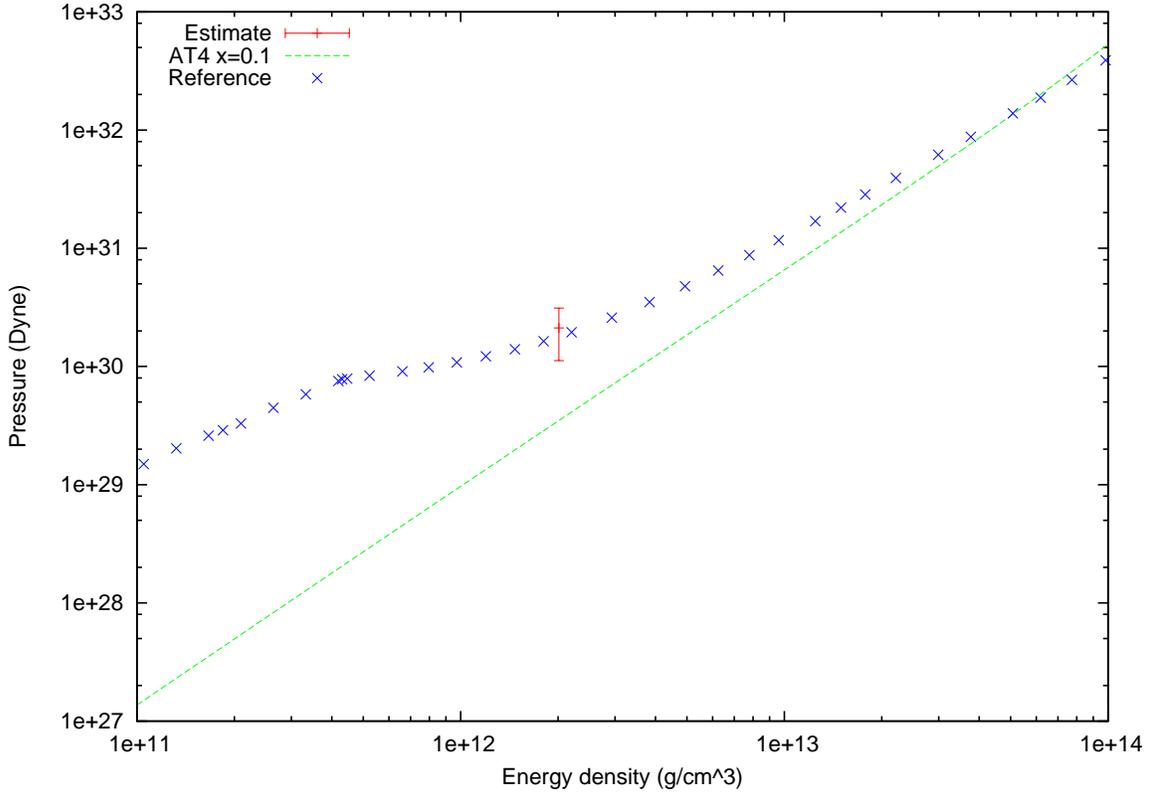}
\caption{Energies ($E$) of admixtures with different $\alpha^2$ at different densities using the AT4' potential.}
\label{fig:eos}
\end{figure}

In figure \ref{fig:eos} we show our EOS at subnuclear density computed in two cases: PNM and 10\% protons. We show in the same plot the BPS EOS \cite{baym1971} as a useful comparison. Quite comfortingly our results show quite a good agreement at the edge of the inner crust. This agreement is obviously not preserved at lower densities due to the absence of clusters. In the same diagram we also show a point computed subtracting, from the energy of the pure gas, the binding energy of the corresponding nucleus (as reported by \cite{Negele1973}) obtained using the semiempirical mass formula. We used different polynomials to fit $E(\rho)$ and hence to derive $P$. The errorbars show our best estimate obtained using a cubic spline interpolation.

\subsection{Particle-hole}

\begin{figure}
\includegraphics[width=0.6\textwidth ,angle=270]{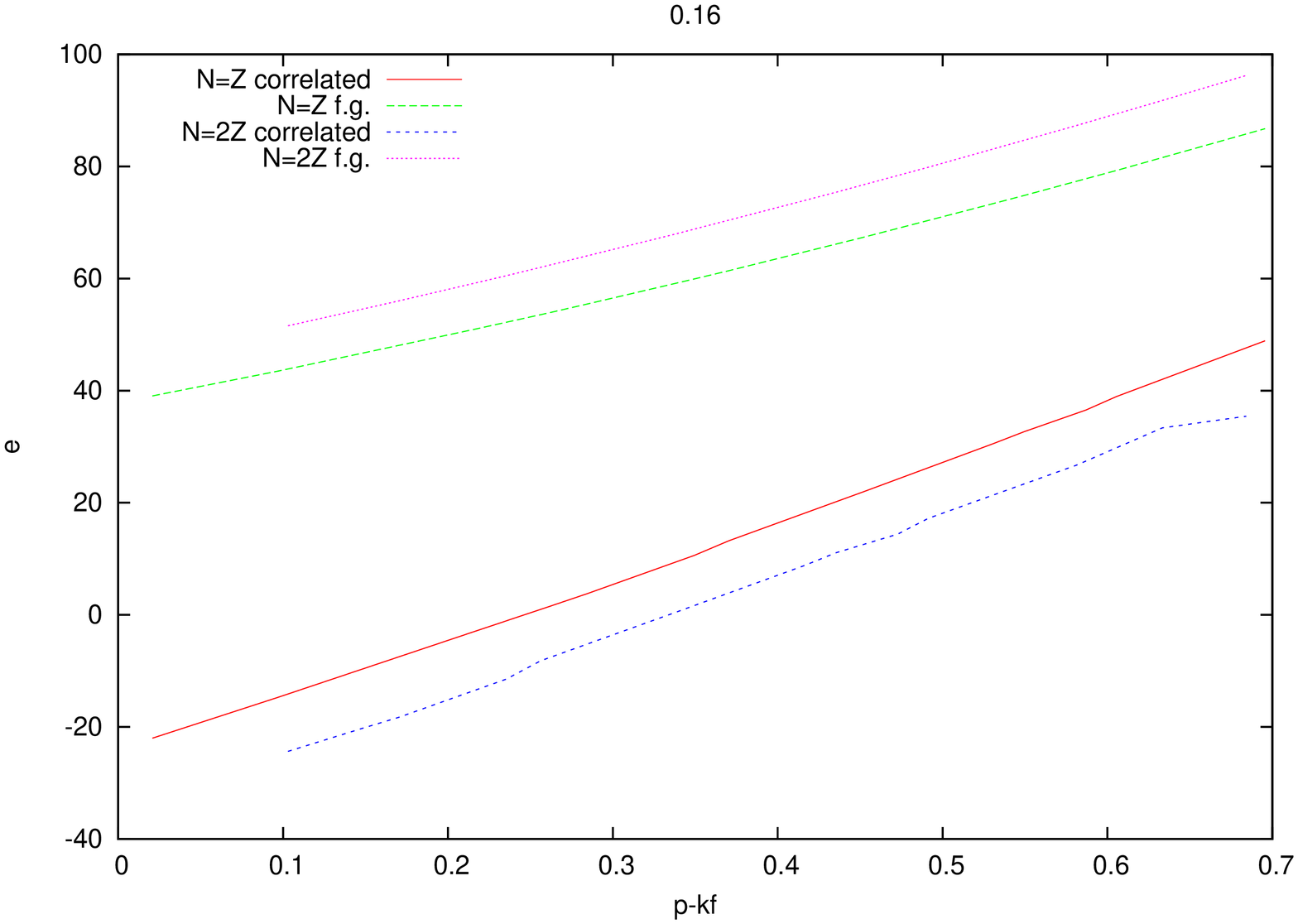}
\includegraphics[width=0.6\textwidth ,angle=270]{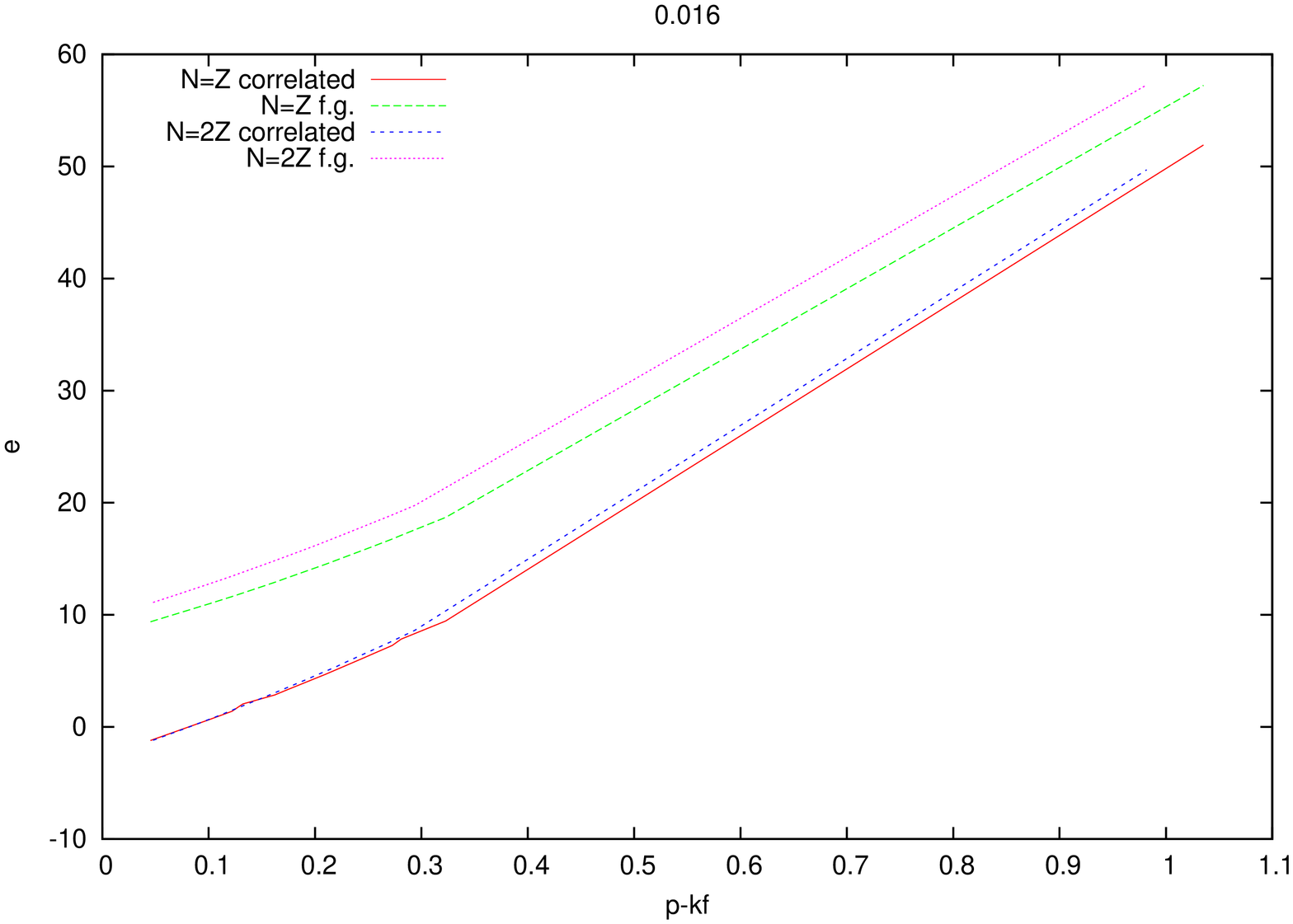}
\caption{Single particle energy $e(p)$ for $N=Z$ and $N=2Z$ at $\rho=\rho_0$ (upper panel) and $\rho=0.1\rho_0$ (lower panel) computed using the AT4' potential. The figures report also the fermi gas estimates. Energies are in MeV and momenta in $\rm{fm^{-1}}$}
\label{fig:ph}
\end{figure}

In fig \ref{fig:ph} we show our results for the single particle energy in different test cases as a function of $q=p-k_f$. The values at $q=0$ are given by the corresponding chemical potentials $\epsilon_{Fa}$ given in eq. \ref{eq:epsfa}. The correlation effects can be viewed by comparing the $\hat{F}_4$ results with the corresponding Fermi gas estimates. Our results at $\rho_0$ and $\alpha=0$ are in reasonably good agreement with the FHNC/SOC calculations of ref \cite{fantoni1983} obtained with the Urbana $V_{14}$ + TNR interaction. One can see that at lower densities the effects on the optical potential due to the asymmetry are much reduced.

\section{Conclusions and perspectives}

We have made a first step towards the development of a technique, based on CBF theory, allowing for a study of the NS crust from first principles. In particular we have developed a theoretical framework suitable for studying NS crustal cells using $\upsilon_4$ potentials and we have applied it, using the AT4' potential as a test case. 

Our results are promising in several respects. First of all it is shown that using only the third component of the isospin dependent correlation is a good enough approximation. We are in a good position, from the variational point of view, to insert a nucleus in our system as discussed in the introduction. We are also in a good position to rewrite the FHNC/SOC scheme, and the CBF perturbative corrections, using this simplified operator to treat the isospin. This would enable us to use more realistic potentials (particularly those with tensor interaction) and hence to refine our results. Notice that standard FHNC/SOC does not allow for asymmetric matter.

Moreover the vertex corrected theory developed here can be readily used to include superfluid effects, thus improving the accuracy of our description of the crustal matter. 

Finally the extension of PBFHNC theory to the treatment of asymmetric matter is essential to deal with the crustal matter, thus, enabling a fully self-consistent description of the neutron star equation of state.

This work was partailly supported by CompStar, a Research Networking Programme of the European Science Foundation. This work was partially funded under MIUR PRIN fund ``Fermi Hypernetted Chain and Quantum Monte Carlo studies of nuclei and nuclear matter with applications to the astrophysics of neutron stars'', and National Science Foundation grand PHY0757703.

\appendix

\section{Computation of correlations \label{app:corr}}

To compute the correlation function needed for our calculation, we have to 
solve a set of differential equations which can be derived by minimizing the 
expectation value of the energy given by the lowest order diagrams, as shown in 
\cite{wiringa1988}. The expectation value of the energy is:
\begin{widetext}
\begin{eqnarray}
E_{LO}&=&\frac{1}{2\rho} \int_{\Omega}d\mathbf{r} \lbrace \frac{\hbar^2}{m} \left( \bigtriangledown f_\parallel \right)^2 \left[ \rho_N^2 \left(1-\frac{1}{2} \ell_N^2 \right) + \rho_P^2 \left(1-\frac{1}{2} \ell_P^2 \right) \right] + \nonumber \\
&+& f_\parallel^2 \left[ \left( \upsilon_c + \upsilon_\tau \right)\left( \rho_N^2 \left(1-\frac{1}{2} \ell_N^2 \right) + \rho_P^2 \left(1-\frac{1}{2} \ell_P^2 \right) \right) -3 \left(\upsilon_\sigma + \upsilon_{\sigma\tau} \right) \left(\frac{1}{2} \rho_P^2 \ell_P^2 + \frac{1}{2} \rho_N^2 \ell_N^2 \right)\right] \rbrace \nonumber \\
&+&  \frac{\rho_P \rho_N}{\rho}\int_{\Omega}d\mathbf{r} \lbrace  \frac{\hbar^2}{m} \left( \bigtriangledown f_\bot \right)^2 + f_\bot^2 \left[  \left( \upsilon_c - \upsilon_{\tau} \right) - 4 \frac{\ell_P}{2} \frac{\ell_N}{2}\left(\upsilon_\tau + 3 \upsilon_{\sigma\tau} \right) \right] \rbrace \, .
\end{eqnarray}
\end{widetext}
Minimizing this expression with respect to $f_\parallel$ and $f_\bot$ gives:
\begin{eqnarray}
&- \frac{\hbar^2}{m}  \left( \bigtriangledown f_\bot \right)^2 + f_\bot \left[\upsilon_c - \upsilon_\tau - \ell_P \ell_nN\left( \upsilon_\tau + 3 \upsilon_{\sigma\tau} \right) \right] =0 \, , \nonumber \\
&- \frac{\hbar^2}{m} \left[  \left( \bigtriangledown f_\parallel \right)^2 G + \bigtriangledown f_\parallel \left( \bigtriangledown G \right)  \right] + f_\parallel \left[ \left(\upsilon_c + \upsilon_\tau \right) G - 3 \left(\upsilon_\sigma + \upsilon_{\sigma \tau}\right) \left( \rho_N^2 + \rho_P^2 - G \right) \right]=0 \, ,
\end{eqnarray}
where:
\begin{equation}
G=\rho_N^2 \left(1-\frac{1}{2} \ell_N^2 \right) + \rho_P^2 \left(1-\frac{1}{2} \ell_P^2 \right)\, .
\end{equation}
 Without any loss of generality, we can divide each term of the 
second equation by $\rho^2$ and redefine G accordingly. In the following, we 
set $\zeta_a=\rho_a / \rho$ .

These two differential equations need to be solved bearing in mind that $f$ 
should heal smoothly to one at some distance $d$. For easily solving them, it 
is convenient to start from $r=0$ after having redefined the variables:
\begin{eqnarray}
\phi_\bot &=& r f_\bot \, ,  \nonumber \\
\psi_\parallel &=& r \sqrt{G} f \, ,\nonumber
\end{eqnarray}
with:
\begin{eqnarray}
\phi_\bot (d) = d && \phi_\bot^\prime(d) = 1 \, ,  \nonumber \\
\psi_\parallel(d) = d \sqrt{G(d)}&& \psi_\parallel^\prime(d)=\sqrt{G(d)}+ \frac{1}{2} \frac{d G^\prime(d)}{\sqrt{G(d)}}  \, .\nonumber
\end{eqnarray}
 These two new functions are defined so that their value at the origin is zero. 
Since there are then three boundary conditions, we need to introduce two 
lagrange multipliers $\lambda$ to ensure that they are all satisfied 
simultaneously. This leads to:
\begin{eqnarray}
&- \frac{\hbar^2}{m}   \phi_\bot^{\prime\prime}+\phi_\bot \left[\upsilon_c - \upsilon_\tau - \ell_p \ell_n \left( \upsilon_\tau + 3 \upsilon_{\sigma\tau} \right)  \right] =\lambda_\phi \phi_\bot \, , \nonumber \\
&- \frac{\hbar^2}{m} \psi_\parallel^{\prime\prime}  + \psi_\parallel \left[\frac{\left(\upsilon_c + \upsilon_\tau \right) G - 3 \left(\upsilon_\sigma + \upsilon_{\sigma \tau}\right) \left( \zeta_n^2 + \zeta_p^2 - G \right)}{G} +  \frac{\hbar^2}{m} \left( \frac{1}{2} \frac{G^{\prime\prime}}{G} + \frac{G^\prime}{rG} - \frac{1}{4} \frac{{G^\prime}^2}{G^2}\right)   \right] = \lambda_\psi  \psi_\parallel \, .
\end{eqnarray}
 We solve these two equations using a standard adaptive-stepsize Bulirsch-Stoer 
method, varying d so as to minimize the energy and iterating to evaluate the 
$\lambda$s. The equations have the form
 \begin{equation}
\phi^{\prime\prime} + \left(a(r) - \lambda \right) \phi =0\, ,
\end{equation}
and we can adjust lambda by adding, at each iteration, a $\delta\lambda$ 
defined as:
\begin{equation}
\delta\lambda = \frac{\phi_T(d)\phi_C^{\prime}(d) - \phi_C(d)\phi_T^{\prime}(d)}{\int_0^d \phi_C^2}\, ,
\end{equation}
 where the quantities denoted with a subscript $C$ are those computed 
numerically at the previous step, and those with a subscript $T$ are 
theoretically derived boundary conditions which need to be satisfied at $r=d$.

\section{Some standard quantum mechanics results \label{app:standardquantum}}

As a useful reference we recall that:
\begin{equation}
\mathcal{A}[\psi_1^{*}\cdots\psi_N^{*}]\hat{O}(x_1,\cdots,x_n)\mathcal{A}[\psi_1\cdots\psi_N]
\end{equation}
can be rewritten as:
\begin{equation}
 \frac{1}{\sqrt{N!}} \sum_{i=1,i \neq j}^{N} [\psi_1^{*}\cdots\psi_N^{*}]\hat{O}(x_1,\cdots,x_n) \mathcal{A}[\psi_1\cdots\psi_N]\, .
\end{equation}
 The anti-symmetrizing operator can be written in terms of the two-particle 
exchange operator $P$:
\begin{equation}
 \mathcal{A}=1 - \sum_{i<j} P_{ij} + \sum_{i<j<k} (P_{ij}P_{jk} + P_{ik}P_{kj}) 
 + \sum_{i<j<k<l} (P_{ij}P_{kl}+P_{ik}P_{jl}+P_{il}P_{jk}) - \{ijkl \,\rm{loop}\} + \cdots \label{eq:exchange}
\end{equation}
 Applying eq. \ref{eq:exchange}, it is easy to check that the fermi-gas n-body 
correlation function is given by

\begin{eqnarray}
 g_n^{FG}(\mathbf{r}_1,\dots,\mathbf{r}_n) &=& 1-\sum_{a,(i<j)}\frac{1}{d} \ell_a(\mathbf{r}_{i,j})^2 + \sum_{a,(i<j<k)} \frac{2}{d^3}
\ell_a(\mathbf{r}_{i,j}) \ell_a(\mathbf{r}_{j,k}) \ell_a(\mathbf{r}_{k,i}) +\dots  \, .
\end{eqnarray}
 
 and, in  a more compact form, by the following determinant
 
 \[
  g_n^{FG}(\mathbf{r}_1,\dots,\mathbf{r}_n) = \left|  \begin{array}{ccccc}
  1   &   \frac{1}{d}\sum_a \ell_a(1,2) &  \frac{1}{d} \sum_a\ell_a(1,3)  &      \dots    &  \frac{1}{d}\sum_a \ell_a(1,n) \\
  \frac{1}{d}\sum_a \ell_a(2,1)  &  1  &   \frac{1}{d}\sum_a \ell_a(2,3)  &      \dots    &  \frac{1}{d}\sum_a \ell_a(2,n) \\
    \dots     & \dots       &  \dots      &  \dots   &  \dots   \\
  \frac{1}{d}\sum_a \ell_a(n,1)  &  \frac{1}{d}\sum_a \ell_a(A,2) &  \frac{1}{d}\sum_a \ell_a(n,3)   &  \dots  &   1
   \end{array}  \right|              
    \]

where $\ell_a(i,j)$ and $\ell_a(\mathbf{r}_{i,j})$ are defined in eq. (\ref{eq:elle}), the summations run over the isospin states, $(a=N,P)$
and $d=2$.  The products of the $\ell_a(i,j)$ operators also implies the matrix elements of the relative spin--isospin states 
(all the spin--isospin states of the particle in a loop must be the same). In infinite matter $\ell_a(\mathbf{r}_{i,j})$ reduces to

\begin{equation}
\ell_a(\mathbf{r}_{i,j}) = \ell(x=k_{Fa} r_{ij}) = \frac{3}{x^3} \left[ \sin(x) - x \cos(x) \right] \, .
\end{equation}
\bibliography{statedeppbfhnc}
\end{document}